\documentclass[prd, 11pt,nofootinbib, superscriptaddress, preprintnumbers,floatfix]{revtex4}
\usepackage{iftex}

\ifPDFTeX
  \usepackage[utf8]{inputenc}
  \usepackage[english]{babel}
\else
  \usepackage{fontspec}
  \usepackage{polyglossia}
  \setmainlanguage[latesthyphen=true]{english}
  \ifXeTeX
    \usepackage{xltxtra}
  \else
    \usepackage{luatextra}
  \fi
\fi

\usepackage{amsmath,amssymb}
\usepackage{graphicx}
\usepackage{subfigure}
\usepackage{hyperref}
\usepackage{nicefrac}
\usepackage{color}
\usepackage{dsfont}

\newcommand{\mpi}{M_{\pi}}
\newcommand{\mpin}{M_{\pi^0}}
\newcommand{\mpic}{M_{\pi^0_c}}

\newcommand{\meta}{M_{\eta_2}}

\newcommand{\tr}{\mathrm{Tr}}
\newcommand{\re}{\mathrm{Re}}
\newcommand{\im}{\mathrm{Im}}

\newcommand{\fig}[1]{Figure~\ref{#1}}
\newcommand{\tab}[1]{Table~\ref{#1}}
\newcommand{\eq}[1]{Eq.~(\ref{#1})}

\makeatletter
\newcommand{\checknextarg}{\@ifnextchar\bgroup{\gobblenextarg}{)}}
\newcommand{\gobblenextarg}[1]{,
  \ref{#1}\@ifnextchar\bgroup{\gobblenextarg}{)}}
\makeatother
\graphicspath{{./}}

\begin{document}

\title{Topological susceptibility and $\eta^\prime$ meson mass\\
from $N_f=2$ lattice QCD at the physical point}
%\preprint{arXiv:XXXX.YYYYY}

\newcommand\imp{Institute of Modern Physics, Chinese Academy of Sciences, 730000 Lanzhou, China}
\newcommand\bn{HISKP and BCTP, Rheinische Friedrich-Wilhelms Universit\"at Bonn, 53115 Bonn, Germany}
\newcommand\rom{Dipartimento di Fisica, Universit{\`a} and INFN di
  Roma Tor Vergata, 00133 Roma, Italy}
\newcommand\romt{Dipartimento di Fisica, Universit{\`a} di Roma Tor Vergata, 00133 Roma, Italy}
\newcommand\bern{Institute for Theoretical Physics, Albert Einstein Center for Fundamental Physics,\\ University of Bern, 3012 Bern, Switzerland}
\newcommand\cypa{Department of Physics, University of Cyprus, PO Box 20537, 1678 Nicosia, Cyprus}
\newcommand\cypb{Computation-based Science and Technology Research Center, The Cyprus Institute}
\newcommand\wup{Fakult\"at f\"ur Mathematik und Naturwissenschaften, Bergische Universit\"at Wuppertal}
\newcommand\roma{Centro Fermi - Museo Storico della Fisica e Centro Studi e Ricerche
Enrico Fermi, Compendio del Viminale, Piazza del
Viminiale 1, I-00184, Rome, Italy}
\newcommand\romb{Dipartimento di Fisica, Universit{\`a} di Roma ``Tor Vergata",
Via della Ricerca Scientifica 1, I-00133 Rome, Italy}
\newcommand\mainzkph{Institut f\"ur Kernphysik, Johann-Joachim-Becher-Weg 45, University of Mainz, 55099 Mainz, Germany}

\author{P.~Dimopoulos}\affiliation{\romt}
\author{C.~Helmes}\affiliation{\bn}
\author{C.~Jost}\affiliation{\bn}
\author{B.~Knippschild}\affiliation{\bn}
\author{B.~Kostrzewa}\affiliation{\bn}
\author{L.~Liu}\affiliation{\imp}
\author{K.~Ottnad}\affiliation{\bn}\affiliation{\mainzkph}
\author{M.~Petschlies}\affiliation{\bn}
\author{C.~Urbach}\affiliation{\bn}
\author{U.~Wenger}\affiliation{\bern}
\author{M.~Werner}\affiliation{\bn}
\collaboration{ETM Collaboration}\noaffiliation

\begin{abstract}
\vspace*{0.3cm}
\begin{center}
 \includegraphics[draft=false,width=.2\linewidth]{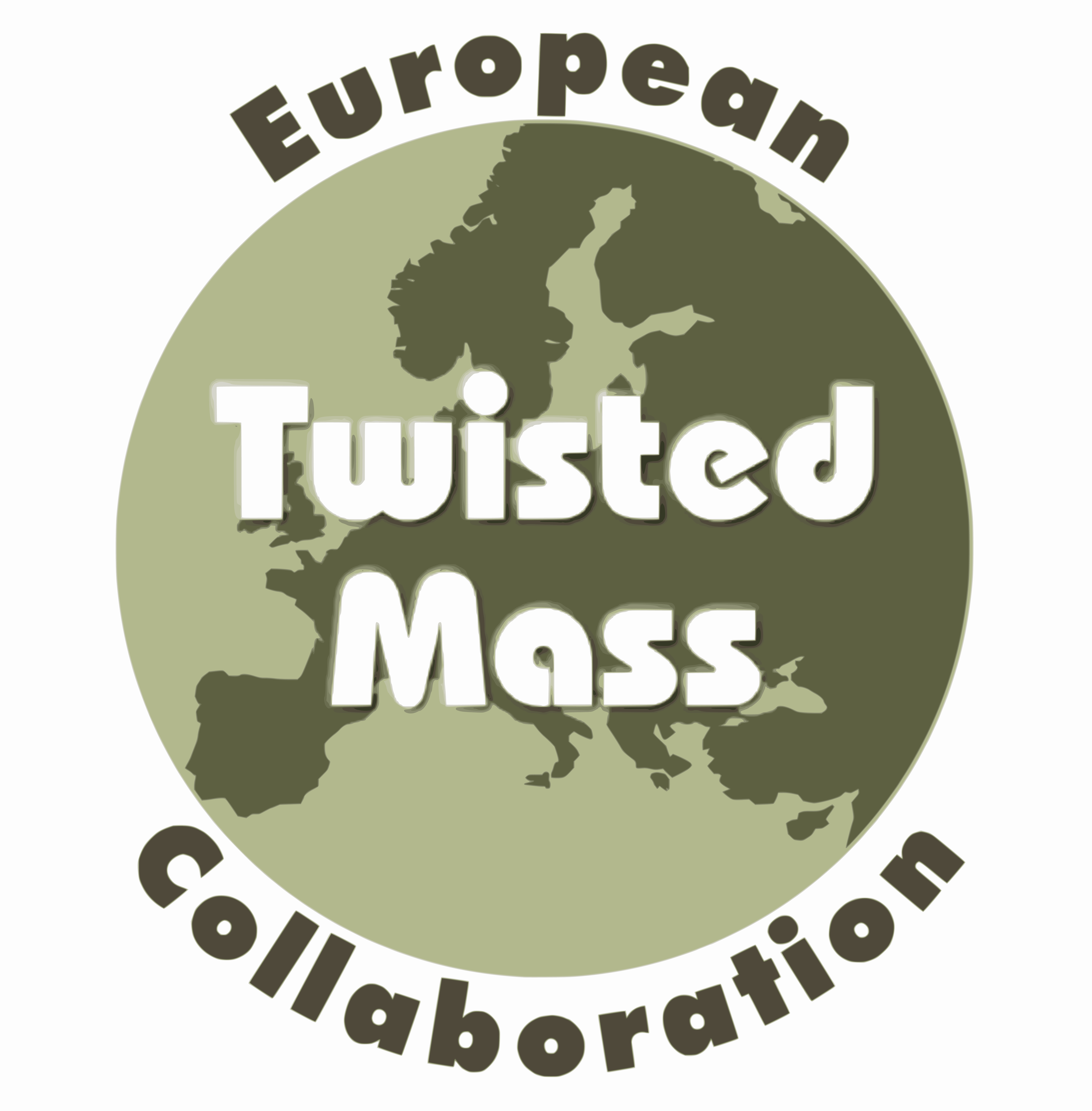}
\end{center}
\vspace*{0.3cm} 
In this paper we explore the computation of topological susceptibility
and $\eta^\prime$ meson mass in $N_f=2$ flavor QCD using lattice
techniques with physical value of the pion mass as well as larger pion
mass values. We observe that the physical point can be reached without a 
significant increase in the statistical noise. The mass of the
$\eta^\prime$ meson can be obtained from both fermionic two point
functions and topological charge density correlation functions, giving
compatible results. With the pion mass dependence of the $\eta^\prime$
mass being flat we arrive at $M_{\eta^\prime}= 772(18)\ \mathrm{MeV}$
without an explicit continuum limit. For the topological susceptibility
we observe a linear dependence on $M_\pi^2$, however, with an
additional constant stemming from lattice artifacts.
\end{abstract}

\maketitle

\clearpage

\section{Introduction}

Due to the persisting $3-5\, \sigma$ deviation in the anomalous
magnetic moment of the muon $a_\mu$ between theory and experiment
there is considerable interest in the decays
$\eta\to\gamma^\star\gamma^\star$ and
$\eta^\prime\to\gamma^\star\gamma^\star$ because a better knowledge
of the corresponding transition form factors could help to reduce the
uncertainty in the hadronic light-by-light contribution to $a_\mu$;
see for instance Ref.~\cite{Jegerlehner:2015stw}.  Moreover, $\eta$
and $\eta^\prime$ mesons are interesting from a theoretical point of
view because the large mass of the $\eta^\prime$ meson is explained by
the anomalously broken $U_A(1)$ axial symmetry in QCD. The
$\eta,\eta^\prime$ mixing pattern and the aforementioned transition
form factors can be computed nonperturbatively using lattice
techniques.

There has been considerable progress in studying $\eta$ and
$\eta^\prime$ mesons from lattice QCD. In
Refs.~\cite{Michael:2013gka,Ottnad:2017bjt} the corresponding mixing
has been studied for three values of the lattice spacing and a large,
but still unphysical range of pion mass values in $N_f=2+1+1$ flavor
QCD. After extrapolation to the physical pion mass value excellent
agreement to experiment was found. Further lattice results for
$\eta,\eta^\prime$ can be found in
Refs.~\cite{Christ:2010dd,Dudek:2011tt,Gregory:2011sg,Dudek:2013yja,Fukaya:2015ara}.

Through the anomaly, the mass of the $\eta^\prime$ is also tightly
connected to topology and in particular the topological susceptibility
$\chi_\text{top}$. The latter quantity must decrease as $M_\pi^2$ toward the
chiral limit, if the $\eta^\prime$ is not a Goldstone
boson~\cite{Leutwyler:1992yt}. For recent lattice studies of the
topological susceptibility; see for
instance~\cite{Aoki:2017paw,Alexandrou:2017bzk}. There is now
particular interest in $\chi_\text{top}$ due to its connection to axion dark
matter; see for instance
Refs.~\cite{Dowrick:1991sj,Moore:2017ond,DiVecchia:2017xpu}.

In this paper we attempt to study the $\eta^\prime$ meson and the
topological susceptibility directly at the physical point, however, in
a first step in $N_f=2$ flavor QCD. In $N_f=2$ flavor QCD there
exist a pion triplet and one flavor singlet, which is related to the
aforementioned anomaly. We will denote it as the $\eta_2$ meson to
distinguish it from the $\eta^\prime$ meson in full QCD, which is only
approximately a flavor eigenstate. The $\eta_2$ and the $\eta^\prime$
meson have in common that both receive significant fermionic
disconnected contributions. In an earlier study~\cite{McNeile:2000hf}
their masses have been found to differ only by $200\ \mathrm{MeV}$,
with the additional strange quark introducing only a moderate shift in
the mass. In particular, both are expected to have a similar
dependence on the light quark mass. The most recent lattice QCD studies
of the $\eta_2$ meson can be found in Refs.~\cite{Sun:2017ipk,Jansen:2008wv}.

We investigate $\meta$ using fermionic correlation functions and in
addition topological charge density correlators. The topological
susceptibility is studied using gradient flow
techniques~\cite{Luscher:2010iy}.  Studying the $\eta_2$ meson and the 
topological susceptibility at the physical point will reveal on the
one hand important qualitative information on the implementation of
the anomaly in QCD. On the other hand it represents a feasibility
study for a later investigation of $\eta$ and $\eta^\prime$ in
$N_f=2+1+1$ QCD at the physical point~\cite{Alexandrou:2018egz}. The
results obtained here are also important prerequisites for an
exploratory study of $\eta_2\to\gamma^\star\gamma^\star$.

The paper is organized as follows: in the following two sections we
discuss the lattice details of our computation. In Sec.~\ref{sec:Analysis Methods} we
present the analysis methods and in Sec.~\ref{sec:Results} the results. We close
with a discussion and summary. For a first account of this work we
refer to Ref.~\cite{Helmes:2017ccf}.

\section{Lattice action}
\label{sec:actions}

\begin{table}[t!]
 \centering
 \begin{tabular*}{.9\textwidth}{@{\extracolsep{\fill}}lccccc}
  \hline\hline
  Ensemble & $\beta$ & $c_{\mathrm{sw}}$ &$a\mu_\ell$  &$(L/a)^3\times T/a$ & $N_\mathrm{conf}$  \\ 
  \hline\hline
  $cA2.09.48$ &2.10 &1.57551 &0.009  &$48^3\times96$ & $615$ \\
  $cA2.30.48$ &2.10 &1.57551 &0.030  &$48^3\times96$ & $352$ \\
  $cA2.30.24$ &2.10 &1.57551 &0.030  &$24^3\times48$ & $352$ \\
  $cA2.60.32$ &2.10 &1.57551 &0.060  &$32^3\times64$ & $337$ \\
  $cA2.60.24$ &2.10 &1.57551 &0.060  &$24^3\times48$ & $424$ \\
  \hline\hline
 \end{tabular*}
 \caption{The gauge ensembles used in this study. The labeling of the ensembles follows the notations in Ref.~\cite{Abdel-Rehim:2015pwa}. In addition to the relevant input parameters we give the lattice volume $(L/a)^3\times T/a$ and the number of evaluated configurations $N_\mathrm{conf}$.}
 \label{tab:setup}
\end{table}

The results presented in this paper are based on the gauge
configurations generated by the ETMC with Wilson clover twisted mass
quark action at maximal twist~\cite{Abdel-Rehim:2015pwa}. We employ
the Iwasaki gauge action~\cite{Iwasaki:1985we}.  The measurements are
performed on a set of $N_f = 2$ ensembles with the pion mass ranging
from its physical value to 340~MeV. In Table~\ref{tab:setup} we list
all the ensembles together with the relevant input parameters, the
lattice volume, and the number of configurations. The lattice spacing
is $a=0.0931(2)\ \mathrm{fm}$ for all five ensembles. More details
about the ensembles are presented in Ref.~\cite{Abdel-Rehim:2015pwa}.

The sea quarks are described by the Wilson clover twisted mass
action. The Dirac operator for the light quark doublet consists of the
Wilson twisted mass Dirac operator~\cite{Frezzotti:2000nk} combined
with the clover term
\begin{equation}
  \label{eq:Dlight}
  D_\ell = D -i\gamma_5\tau_3 \left[W_\mathrm{cr} + 
  \frac{i}{4}c_\mathrm{sw}\sigma^{\mu\nu}\mathcal{F}^{\mu\nu}\right] + \mu_\ell\,,
\end{equation}
which acts on a flavor doublet spinor $\psi = (u,d)^T$. In
Eq.~(\ref{eq:Dlight}) we have $D =
\gamma_\mu(\nabla^\ast_\mu+\nabla_\mu)/2$ with $\nabla_\mu$ and
$\nabla_\mu^\ast$ the forward and backward lattice covariant
derivatives, and the Wilson term $W_\mathrm{cr} =
-ra\nabla_\mu^\ast\nabla_\mu + m_\mathrm{cr}$ with the critical mass
$m_\mathrm{cr}$, the Wilson parameter $r=1$, and the lattice spacing
$a$.  The average up/down (twisted) quark mass is denoted by
$\mu_\ell$, while $c_\mathrm{sw}$ is the so-called
Sheikoleslami-Wohlert improvement coefficient
\cite{Sheikholeslami:1985ij} multiplying the clover term. It is in our
case not used for $\mathcal{O}(a)$ improvement but serves to
significantly reduce the effects of isospin
breaking~\cite{Abdel-Rehim:2015pwa}.

The critical mass has been determined as described in
Refs.~\cite{Chiarappa:2006ae,Baron:2010bv}. This guarantees automatic
$\mathcal{O}\left(a\right)$ improvement \cite{Frezzotti:2003ni}, which
is one of the main advantages of the Wilson twisted mass formulation
of lattice QCD.

\section{Observables}

As a smearing scheme in the computation of fermionic correlation
functions we use the stochastic Laplacian Heaviside (sLapH)
method~\cite{Peardon:2009gh,Morningstar:2011ka}.  The details of our
sLapH parameter choices for a set of $N_f = 2+1+1$ Wilson twisted mass
ensembles are given in Ref.~\cite{Helmes:2015gla}. The parameters for
the ensembles used in this work are the same as those for $N_f =
2+1+1$ ensembles with the corresponding lattice volume.

\subsection{\texorpdfstring{$\eta_2$}{eta2} and pion correlation functions}

In $N_f=2$ flavor QCD there is the neutral pion, corresponding to the
neutral of the three pions in the triplet, and the $\eta_2$, the
flavor singlet pseudoscalar meson related to the axial $U_A(1)$
anomaly. Since up and down quarks are mass degenerate, there is no
mixing among the neutral pion and the $\eta_2$ with our action. We
employ the following pseudoscalar interpolating operators projected to
zero momentum, which are all local and Hermitian
\begin{equation}
  \mathcal{P}^3(t)\ =\ \frac{1}{\sqrt{2}}\sum_\mathbf{x} \bar\psi
  i\gamma_5\,\tau^3\, \psi(\mathbf{x},t)\,,\qquad
  \mathcal{P}^0(t)\ =\ \frac{1}{\sqrt{2}}\sum_\mathbf{x} \bar\psi
  i\gamma_5\,\mathds{1}_f\, \psi(\mathbf{x},t)\,.
\end{equation}
Here, $\tau^3$ is the third Pauli and $\mathds{1}_f$ the unit matrix,
both acting in flavor space. From those one builds the correlation
functions
\begin{align}
 C_{\pi^0}(t-t')\ &=\ \langle \mathcal{P}^3(t)\,  (\mathcal{P}^3(t'))^\dagger \rangle\,, \label{eq:corr_pi0} \\
 C_{\eta_2}(t-t')\ &=\ \langle \mathcal{P}^0(t)\, (\mathcal{P}^0(t'))^\dagger \rangle\,, \label{eq:corr_eta2}
\end{align}
which allow one to determine the masses $\mpin$ and $\meta$ from their
decay in Euclidean time. Both correlation functions in Eqs.~(\ref{eq:corr_pi0}) 
and~(\ref{eq:corr_eta2}) do have a fermionic connected and a
fermionic disconnected contribution, the latter of which vanishes
exactly in case of the neutral pion in an isospin symmetric
theory. Since this is not the case for Wilson twisted mass fermions,
we have to take the disconnected contributions into account also for
the $\pi^0$.

For the disconnected part of $C_{\eta_2}$ we consider the loop
\begin{equation}
  \label{eq:disc}
  \begin{split}
    \langle\bar\psi_u i\gamma_5\psi_u(x) + \bar\psi_d i\gamma_5\psi_d(x)
    \rangle_F &= -i\,\tr\{\gamma_5G_u^{xx}\} - i\,\tr\{\gamma_5G_d^{xx}\}\\
    &= -i\,\tr\{\gamma_5G_u^{xx}\} - i\,\tr\{(G_u^{xx})^\dagger\gamma_5\}\\
    &= -2i\,\re\tr\{\gamma_5G_u^{xx}\}\,.\\
  \end{split}
\end{equation}
Here, we have used the $\gamma_5$ hermiticity property $D_d = \gamma_5
D_u^\dagger\gamma_5$. $G_{u/d}^{xy}$ represents the \emph{up} or
\emph{down} propagator. Similarly, one shows for $C_{\pi^0}$
\begin{equation}
 \label{eq:disc2}
  \langle\bar\psi_u i\gamma_5\psi_u(x) - \bar\psi_d i\gamma_5\psi_d(x)
    \rangle_F \ =\ 2\,\im\tr\{\gamma_5G_u^{xx}\}\,.
\end{equation}
The fermionic connected contribution is identical for the two
correlation functions \eq{eq:corr_pi0} and \eq{eq:corr_eta2}. With
similar arguments as for the loops one finds
\begin{equation}
  \label{eq:conn}
  C^\mathrm{conn}(t-t')\ =\ \re\tr\{\gamma_5 G_u^{tt'}\gamma_5 G_u^{t't}\}\,,
\end{equation}
where we have suppressed the spatial indices.  From Eqs.~(\ref{eq:disc}),~(\ref{eq:disc2}), and~(\ref{eq:conn}) 
we infer the expressions for the $\pi^0$ and
$\eta_2$ correlation functions as follows:
\begin{equation}
  \label{eq:trcorr}
  \begin{split}
    C_{\pi^0}(t-t')\ &=\ \tr\{\gamma_5 G_u^{tt'}\gamma_5 G_u^{t't}\}
    +2\,\im\tr\{\gamma_5G_u^{tt}\} \cdot\im\tr\{\gamma_5G_u^{t't'}\}\,,\\
    C_{\eta_2}(t-t')\ &=\ \tr\{\gamma_5 G_u^{tt'}\gamma_5 G_u^{t't}\}
    -2\,\re\tr\{\gamma_5G_u^{tt}\} \cdot\re\tr\{\gamma_5G_u^{t't'}\}\,.\\
  \end{split}  
\end{equation}
For completeness, the correlation function of the charged pion is
constructed as
\begin{equation}
  C_{\pi^\pm}(t-t')\ =\ \langle \mathcal{P}^+(t)\ (\mathcal{P}^+(t'))^\dagger\rangle
\end{equation}
with
\begin{equation}
  \mathcal{P}^+(t)\ =\ \sum_\mathbf{x} \bar\psi i\gamma_5 \frac{\tau^1
     + i\tau^2 }{2}\psi(\mathbf{x},t)
\end{equation}
and $\tau^1$ and $\tau^2$ the first and second Pauli matrices, respectively.

\subsection{Topological charge density correlations and
  susceptibility}
\label{subsec:action_topcharge}
The naive field theoretical definition of the topological charge
density given by
\begin{equation}
q(x) = - \frac{1}{32\pi^2} \varepsilon_{\mu\nu\rho\sigma} \tr
F_{\mu\nu}(x) F_{\rho\sigma}(x)
\end{equation}
defines the topological charge density point-to-point correlator
\begin{equation}
\label{eq:definition_charge_correlator}
C_{qq}(x-y) = \langle q(x) q(y)\rangle \, .
\end{equation}
The topological susceptibility, which is a measure for the
fluctuations of the topological charge, is defined as
\begin{equation}
  \label{eq:topsusdef}
  \chi_\text{top}\ =\ \frac{1}{V}\int \mathrm{d}x \int \mathrm{d}y\ \langle
  q(x)\, q(y)\rangle
\end{equation}
where $V$ is the spacetime volume.  Due to the pseudoscalar nature of
the topological charge density the topological charge density
correlator is strictly negative for finite separations, $C_{qq}(x-y>0)
< 0$. On the other hand, it is clear that the susceptibility is
strictly positive, because $V \cdot \chi_\text{top} = \langle
Q^2\rangle > 0$, where $Q=\int dx \, q(x)$ is the total topological
charge of the gauge field. This apparent contradiction is resolved by
recalling that $C_{qq}$ suffers from contact-term singularities at
$x-y=0$ which need to be renormalized in order for the susceptibility
to make physical sense. Hence, the physics of the topological
susceptibility is intricately hidden in the difference between the
contact-term contribution of the correlator at $|x-y|=0$ and the
contributions at $|x-y| > 0$.

Another interesting property of the topological charge density
correlator is that it couples to the flavor singlet pseudoscalar
mesons. It is in fact this coupling which is thought to be responsible
for the large mass of the $\eta_2$. As a consequence, the behavior of
the topological charge density point-to-point correlator is dominated
by the single boson propagator for the $\eta_2$ meson and therefore
follows the form of the scalar propagator
\cite{Dowrick:1991sj,Shuryak:1994rr}
\begin{equation}
  C_{qq}(x,y) \sim \frac{M}{|x-y|} K_1(M \cdot|x-y|)
  \label{eq:asymptotic_scalar_propagator}
\end{equation}
where $K_1$ is the modified Bessel function of the first kind, and $M$
is the mass of the lightest particle in the pseudoscalar meson sector,
i.e., the mass $M_{\eta_2}$ of the $\eta_2$ meson.

On the lattice the topological charge density is discretized with a
clover-type discretization of the field strength tensor $F_{\mu\nu}$
which extends over a distance of $2a$ in lattice units. Hence the
contact-term contributions to the topological charge density
correlator are also distributed over the distance $|x-y| \sim
2a$. Moreover, since the discretized field strength tensor is
constructed from smeared gauge links in order to remove ultraviolet
fluctuations of the gauge field, the positive contact-term
contributions to the correlator are spread over a range $R_0$ which
depends on the details of the smoothing scheme. However, the behavior
of the correlator for $|x-y| \gg R_0$ should be independent of the
details of the smearing scheme and hence any smoothing scheme is
supposed to yield the same physics, i.e.~the same mass $M_{\eta_2}$.

For the topological charge density correlator we use the array processor experiment (APE) smearing
scheme~\cite{Albanese:1987ds} with various smearing levels ranging up
to 90 iterative smearing steps. This is in order to check for the
independence of the results from the smearing scheme. To compute the
topological susceptibility $\chi_\text{top}$ we employ the gradient
flow technique as introduced for lattice QCD in
Ref.~\cite{Luscher:2010iy}. It has the advantage of yielding a
renormalized topological susceptibility at finite flow time
$t$~\cite{Luscher:2011bx}, in particular it renormalizes the contact
term singularities in the continuum limit at any fixed, physical value
of $t$. Since the renormalized susceptibility is scale invariant,
i.e., independent of the renormalization scale, $\chi_\text{top}$
becomes independent of the flow time $t$ at sufficiently large $t$
toward the continuum limit. This is indeed what we observe in our
calculation. However, we note that lattice artifacts might well be
very different for the susceptibility at different values of
$t$. Instead of calculating the topological susceptibility via the
lattice version of \eq{eq:topsusdef}, we first obtain the topological
charge at flow time $t$ from the topological charge density $q_t(x)$
evaluated on the flown gauge field configuration,
\begin{equation}
Q(t) = a^4 \sum_x q_t(x) \, ,
\end{equation}
and then the susceptibility via $\chi_\text{top}(t) = \langle
Q(t)^2\rangle/V$.  We choose $t=3t_0$ where $t_0$ is the usual
gradient flow reference scale defined through the renormalized action
density~\cite{Luscher:2010iy} on the corresponding ensemble. In
addition, we also make use of the related reference scale $t_1$ in
order to facilitate the comparison of our results with those in
Ref.~\cite{Bruno:2014ova}.

\section{Analysis Method}
\label{sec:Analysis Methods}

In the following Secs.~\ref{subsec:Excited state subtraction} and~\ref{subsec:Shifted correlation functions} we will
first give more details on the analysis of the fermionic correlation
function, before we turn to the discussion for the gluonic correlators
in Sec.~\ref{subsec:Charge density correlators}. \par

The fermionic correlation function data are generally analyzed using
the blocked bootstrap procedure with $R=10000$ bootstrap
samples. Depending on the ensemble, we have chosen the block size such
that at least $\gtrsim 100$ blocked data points are left. The relevant
masses are computed from correlated fits to the correlation function
data.  \par

The gluonic correlation function data are analyzed using a jackknife
procedure. It turns out that the correlators at separate distances are
highly correlated even for large separations, such that the covariance
matrix cannot be taken into account reliably in the fitting procedur;e
see further details below.  The data for the topological charge
susceptibility (and the gluonic scales $t_0$ and $t_1$) are analyzed
using a blocked bootstrap procedure with $R=1000$ bootstrap samples
and block sizes such that $\gtrsim 30$ blocked data points are
left. The so obtained error is compared to the naive one corrected by
the integrated autocorrelation time $\tau_\text{int}$, and the larger
of the two is always chosen as the final error. Since these
calculations are inexpensive, we use at least double the number of
configurations indicated in Table \ref{tab:setup}. \par

\subsection{Excited state subtraction}
\label{subsec:Excited state subtraction}

In particular for the $\eta_2$ meson, the fermionic disconnected
contributions are very noisy. As a consequence, the signal is lost
relatively early in Euclidean time. For this reason we have in the
past applied a method to subtract excited
states~\cite{Jansen:2008wv,Michael:2013gka,Liu:2016cba,Ottnad:2017bjt},
originally proposed in Ref.~~\cite{Neff:2001zr}. It actually works
very well and we will apply it here again for the $\eta_2$ meson. It
consists of subtracting excited states from the connected contribution
only. This is feasible, because the connected part --- representing a
pion correlation function --- has a signal for all Euclidean time
values. Therefore, we can fit to it at large enough Euclidean times
such that excited states have decayed sufficiently. Next, we replace
the connected correlation function at small times by the fitted
(ground state) function. Thereafter, the so subtracted connected
contribution is summed according to \eq{eq:trcorr} to the full
$\eta_2$ correlation functions.

The underlying assumption is that disconnected contributions are large
for the ground state, i.e. the $\eta_2$, but not for excited
states. If this assumption is correct, the effective mass
\begin{equation}
  M_\mathrm{eff}\ =\ -\log\frac{C_{\eta_2}(t)}{C_{\eta_2}(t+1)}
\end{equation}
should show a plateau from very early Euclidean times on. We have
found in Refs.~\cite{Jansen:2008wv,Michael:2013gka,Ottnad:2017bjt}
that this approach works very well for the $\eta_2$ meson in $N_f=2$
flavor QCD as well as for $\eta$ and $\eta^\prime$ mesons in
$N_f=2+1+1$ flavor QCD.

\subsection{Shifted correlation functions}
\label{subsec:Shifted correlation functions}

The expected time dependence of the fermionic correlation functions
considered here reads as follows:
\begin{equation}
  C(t) = |\langle 0|\mathcal{O}|0\rangle|^2 + \sum_n \frac{|\langle 0 |
  \mathcal{O}| n\rangle|^2}{2E_n}\left(e^{-E_n t} + e^{-E_n
    (T-t)}\right)\,,  
\end{equation}
where $\mathcal{O}=\mathcal{P}^+,\mathcal{P}^3,\mathcal{P}^0$ and $n$
labels the states with the corresponding quantum numbers. The time
independent first term on the right-hand side corresponds to the
vacuum expectation value (VEV). Using the symmetries of our action one
can show that for $\mathcal{P}^+$ and $\mathcal{P}^0$ the VEV must be
zero, while this is not the case for $\mathcal{P}^3$. We deal with the
VEV by building the shifted correlation function
\begin{equation}
  \label{eq:corrshifted}
  \tilde{C}(t)\ =\ C(t) - C(t+1)\,.
\end{equation}
The difference cancels the constant VEV contribution, while also
changing the time dependence to be antisymmetric in time,
\begin{equation}
  \tilde{C}(t)\ \propto\ \left(e^{-E_n t} - e^{-E_n (T-t)}\right)\,.
\end{equation}
As an alternative, one can also compute the VEV $|\langle
0|\mathcal{O}|0\rangle|^2$ from the data and subtract it
explicitly.

Since the VEV has to be zero for $\mathcal{P}^0$ up to statistical
fluctuations, strictly speaking we do not need to use the shifting
procedure for the $\eta_2$ meson. However, as has been argued in
Ref.~\cite{Aoki:2007ka} and first investigated in
Ref.~\cite{Bali:2014pva}, there is an additive finite volume effect to
$C_{\eta_2}$ constant in Euclidean time of the form
\begin{equation}
  \propto\ \frac{a^5}{T}\left(\chi_\text{top} + \frac{Q^2}{V}\right)
\end{equation}
proportional to the topological susceptibility $\chi_\text{top}$ and the
squared topological charge $Q^2$.  If present, such a term will cause
the $\eta_2$ correlation function to stay finite at large Euclidean
times. Depending on the sign of the coefficient in front of the finite
volume effect, the correlation function may even turn negative at
relatively small Euclidean times.  Clearly, a finite volume effect of
this type can be subtracted again using the shifting procedure, which
has first been proposed and applied in Ref.~\cite{Ottnad:2017bjt}.

\subsection{Topological charge density correlators}
\label{subsec:Charge density correlators}

For the computation of the topological charge density correlator we
make use of the full translational invariance. In order to do so, we
obtain the topological charge density correlator in
\eq{eq:definition_charge_correlator} by Fourier transforming the
topological charge density on each gauge field configuration,
calculating the correlator in Fourier space and transforming it back
to coordinate space.  In this sense the evaluation is exact, in
contrast to the computation of the disconnected contributions to the
fermionic correlators in Eqs.~(\ref{eq:corr_pi0}) and~(\ref{eq:corr_eta2}), which can
only be evaluated stochastically.

The employed smearing level has several effects on the correlation
function $C_{qq}(x-y)$. First, it reduces the statistical errors
because the smearing suppresses ultraviolet fluctuations. Hence, with
increasing smearing levels the signal can be followed over larger and
larger separations $x-y$. Second, the increased smearing enhances
the contribution of the ground state in the correlation function,
i.e.~in this case the contribution of the $\eta_2$ state. Third,
with an increasing smearing level the contact term is distributed over
larger distances and hence distorts the correlation function up to
larger and larger separations. Obviously, these effects compete with
each other with respect to the optimal fit range.

\begin{figure}[thb] 
\includegraphics[height=.465\linewidth]{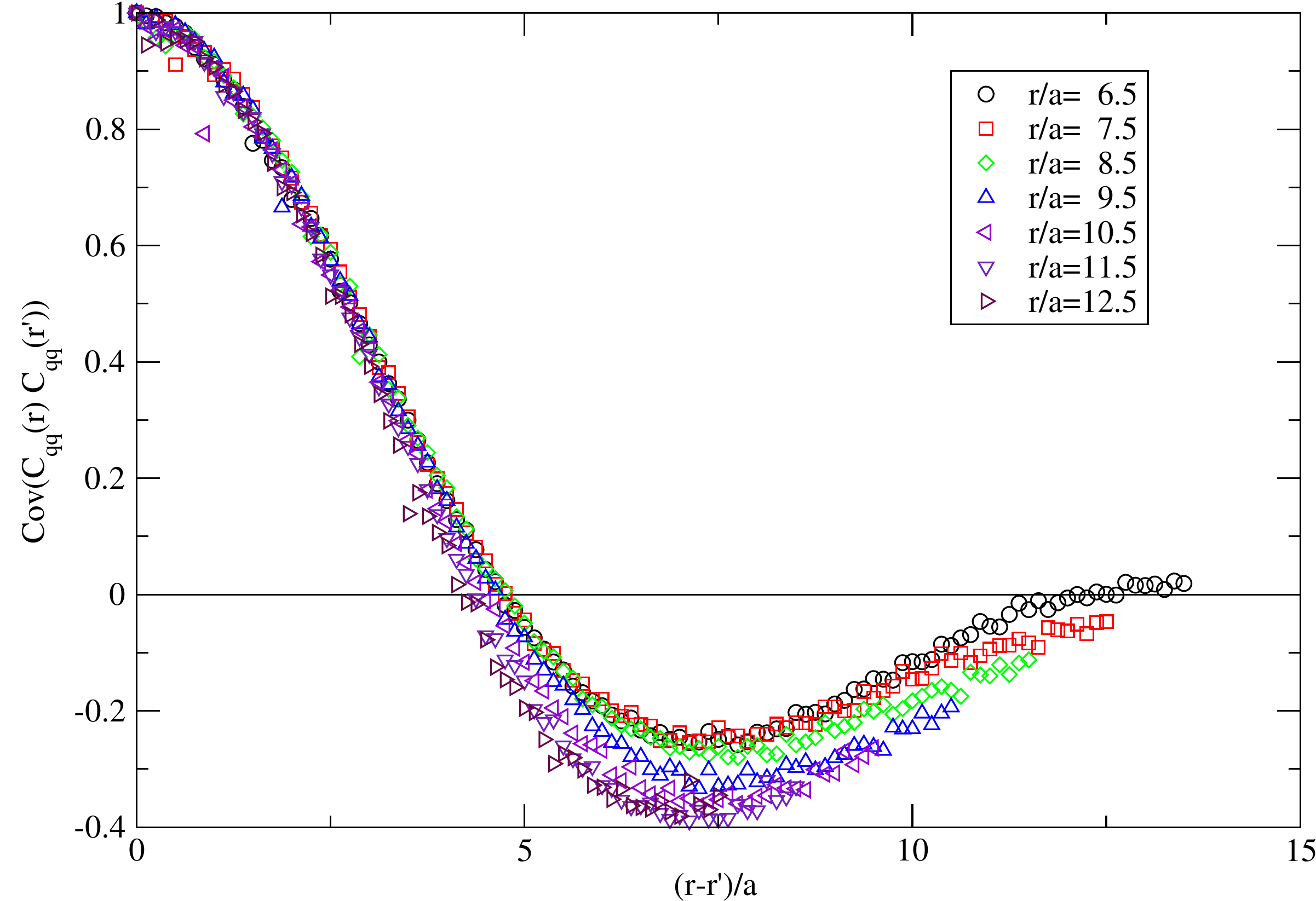}
\caption{Covariance of the binned correlation functions as a function
  of $(r-r')/a$ for different values of $r$ and smearing level $n=90$
  on ensemble cA2.09.48. }
  \label{fig:ffd_covariance_APE90}
\end{figure}

In principle, the choice of the fit ranges should be determined by the
quality of the fits. Unfortunately, here this is not possible, because
the correlators at separate distances $r$ and $r'$ are highly
correlated. We illustrate this in \fig{fig:ffd_covariance_APE90} where
we show the covariance Cov($C_{qq}(r),C_{qq}(r')$) of the correlation
functions as a function of $(r-r')/a$ for different values of $r$ and
smearing level $n=90$.\footnote{The data for the other smearing levels
  look very similar.}
We are essentially looking at separate columns of the covariance
matrix. For ease of comparison we normalize the covariance by
Cov($C_{qq}(r),C_{qq}(r)$), and in addition bin the data into bins of
size $\Delta r/a = 0.125$.

We see that the $C_{qq}$'s are positively correlated for $(r'-r)/a
\lesssim 4.5$ and become more and more strongly anticorrelated until a
maximum of anticorrelation is reached at around $(r'-r)/a \sim
7.5$. Since this correlation is essentially independent of $r$, the
columns of the covariance matrix are highly linear dependent and the
matrix itself is very ill-conditioned. As a consequence, it cannot be
taken into account for reliably estimating the quality of the
$\chi^2$-fits.

We note that all the above conclusions hold independently of the bin
size and the smearing level, and we suspect that the peculiar
behavior is due to some underlying structure in the topology of the
gauge fields.

\section{Results}
\label{sec:Results}
\begin{table}[t!]
 \centering
 \begin{tabular*}{\textwidth}{@{\extracolsep{\fill}}lccccccc}
  \hline\hline
  Ensemble     & $aM_{\pi}$ & $a\mpin$     & $a\mpic$     & $a\meta^\text{ferm.}$ & $a\meta^\text{gl.}$
  & $t_1/a^2$ &  $10^3 \cdot t_1^2 \chi_\text{top}$ \\
  \hline
%   $cA2.09.48$  & $0.06211(06)$  & $0.0576(25)$ & $0.1196(02)$ & $0.361(14)$ & $0.53(4)$\\
%   $cA2.30.48$  & $0.11199(06)$  & $0.0976(35)$ & $0.1521(01)$ & $0.376(11)$ & $0.63(6)$\\
%   $cA2.30.24$  & $0.11461(37)$  & $0.1110(95)$ & $0.1519(29)$ & $0.425(22)$ & $0.52(6)$\\
%   $cA2.60.32$  & $0.15783(12)$  & $0.1555(64)$ & $0.1883(03)$ & $0.396(10)$ & $0.91(5)$\\
%   $cA2.60.24$  & $0.15908(28)$  & $0.1347(86)$ & $0.1883(11)$ & $0.399(12)$ & $0.91(8)$\\
  $cA2.09.48$  & $0.06211(06)$  & $0.0576(25)$ & $0.1196(02)$ & $0.361(14)$ & $0.369(10)$ & $6.890(08)$ & $0.48(3)$\\
  $cA2.30.48$  & $0.11199(06)$  & $0.0976(35)$ & $0.1521(01)$ & $0.376(11)$ & $0.356(17)$ & $6.761(08)$ & $0.56(5)$\\
  $cA2.30.24$  & $0.11461(37)$  & $0.1110(95)$ & $0.1519(29)$ & $0.425(22)$ & $0.386(45)$ & $6.828(30)$ & $0.46(4)$\\
  $cA2.60.32$  & $0.15783(12)$  & $0.1555(64)$ & $0.1883(03)$ & $0.396(10)$ & $0.379(14)$ & $6.562(08)$ & $0.78(4)$\\
  $cA2.60.24$  & $0.15908(28)$  & $0.1347(86)$ & $0.1883(11)$ & $0.399(12)$ & $0.345(48)$ & $6.550(17)$ & $0.79(7)$\\
  \hline\hline
 \end{tabular*}
 \caption{Results for the masses of the charged and the neutral pion
   (full and quark-connected only), the $\eta_2$ meson in lattice
   units (from fermionic and gluonic correlators), the
   gluonic gradient flow lattice scale $t_1/a^2$, and the topological susceptibility in units of $t_1$ for the five ensembles considered.} 
 \label{tab:results}
\end{table}

In \tab{tab:results} we show results for the pion and
$\eta_2^\text{ferm.}$ masses which have been computed from fermionic
correlation functions, the $\eta_2^\text{gl.}$ masses obtained from
the gluonic topological charge density correlation functions, as well
as the gluonic gradient flow lattice scale $t_1/a^2$ and the
topological susceptibility $\chi_\text{top}$. For the charged pion we will
always use the shorthand $M_\pi$, while for the neutral pion $\mpin$
and $\mpic$ refer to the full and quark-connected masses,
respectively. In the following we discuss these results in more
detail.

\begin{figure}[t]
  \centering
  \subfigure{\includegraphics[width=.49\linewidth]{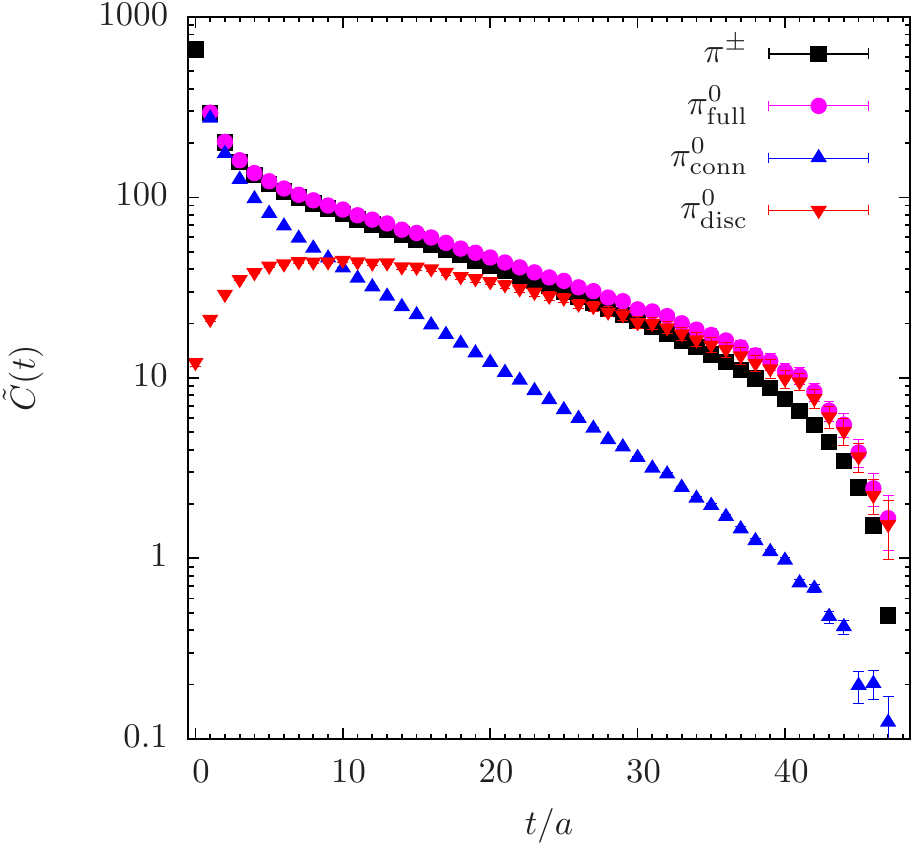}}
  \subfigure{\includegraphics[width=.49\linewidth]{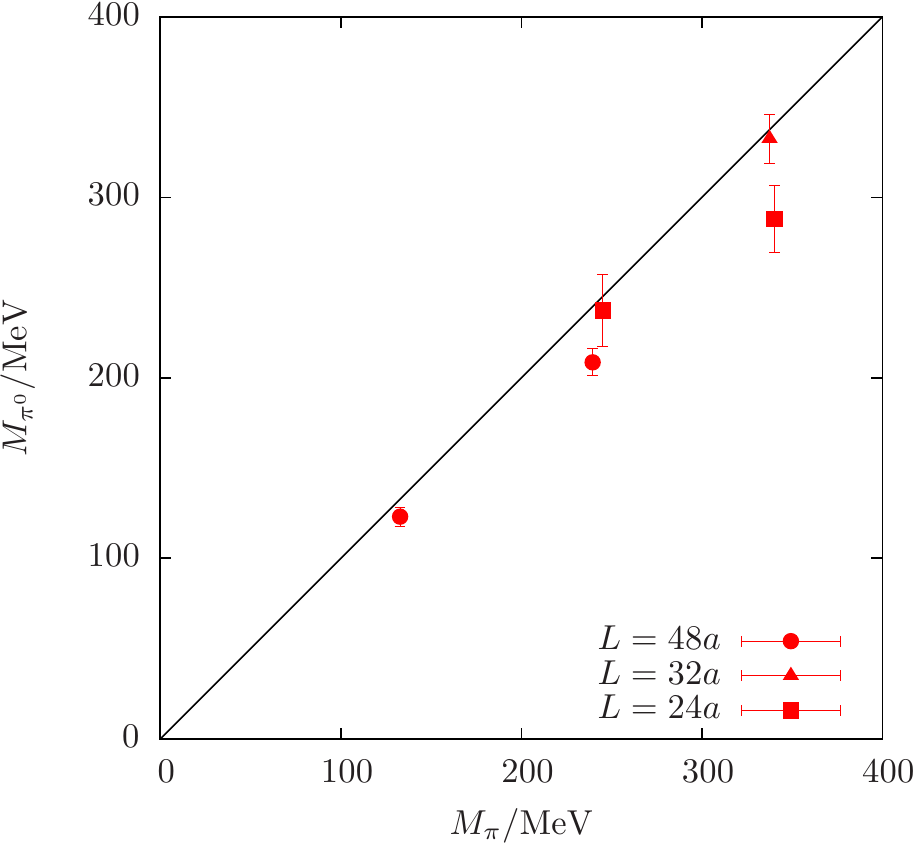}}
  \caption{Overview of results for the pions. Left panel: Shifted correlation functions $\tilde{C}(t)$ for the charged and the neutral pion on ensemble cA2.09.48 with physical quark mass. In the case of the neutral pion we show the full correlation function as well as individual quark-connected and quark-disconnected contributions. Right panel: Mass of the (full) neutral pion as a function of the charged pion mass.}
  \label{fig:pisummary}
\end{figure}

\subsection{Neutral pion}
In contrast to the $\eta_2$ meson discussed later, the signal for the
neutral pion can be resolved for all values of $t/a$. In the left
panel of \fig{fig:pisummary} we show the shifted correlation function
$\tilde{C}(t)$ for the neutral pion as well as the individual
quark-connected and quark-disconnected contributions. Note that in
this case the function shift is required to remove the offset from
the vacuum expectation value in the quark-disconnected
contribution. Clearly the signal in the quark-disconnected part is
well behaved even for the largest values of $t/a$. For comparison the
charged pion has been included in the plot as well. \par 

As already visible from the left panel of \fig{fig:pisummary}, charged
and neutral pions appear to have very similar mass values. This is
even more apparent from the right panel where the neutral pion mass
values are plotted versus the charged pion mass values, both in
physical units, for all the ensembles considered here. The points fall
almost on the bisecting line, which indicates no mass splitting
between neutral and charged pion mass. This finding, which we pointed
out already in Ref.~\cite{Abdel-Rehim:2015pwa}, is rather important:
this mass splitting is basically the only large $a^2$ lattice artefact
that was found for simulations with Wilson twisted fermions at maximal
twist (see also
Refs.~\cite{Jansen:2005cg,Dimopoulos:2009qv}). Including the clover
term appears to reduce its size drastically. We refer to
Ref.~\cite{Herdoiza:2013sla} for a systematic investigation of this
splitting for simulations without the clover term.

\subsection{\texorpdfstring{$\eta_2$}{eta2} meson mass from fermionic correlators}
\label{subsec:Results meson mass from fermionic correlators}

\begin{figure}[t]
  \centering
  \subfigure{\includegraphics[width=.49\linewidth]{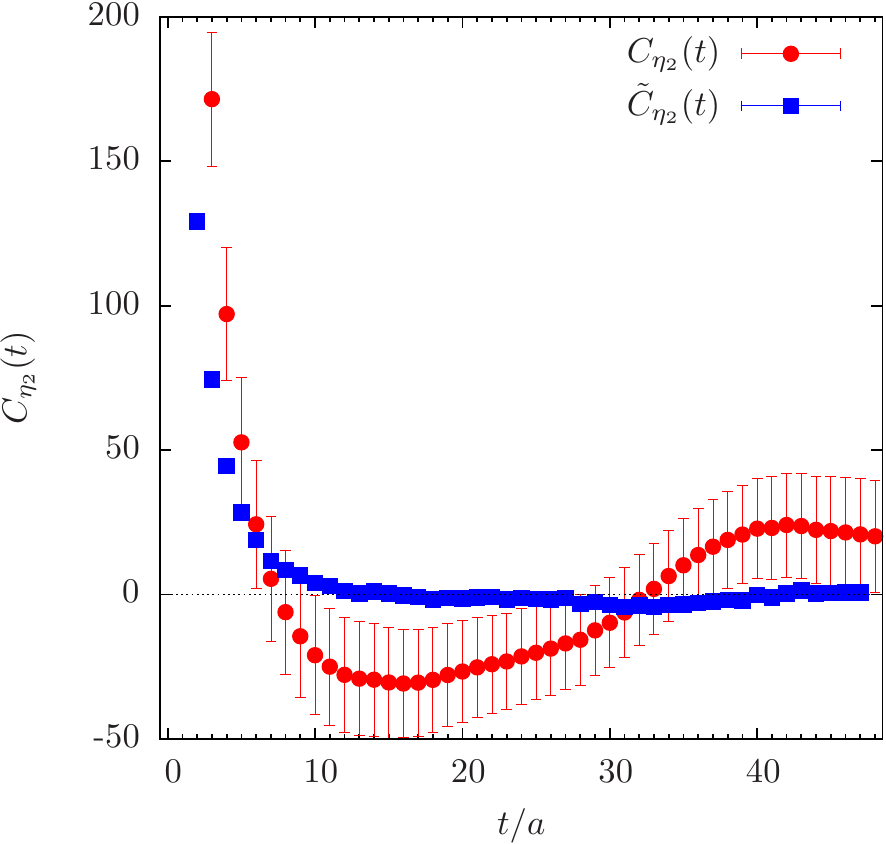}}
  \subfigure{\includegraphics[width=.49\linewidth]{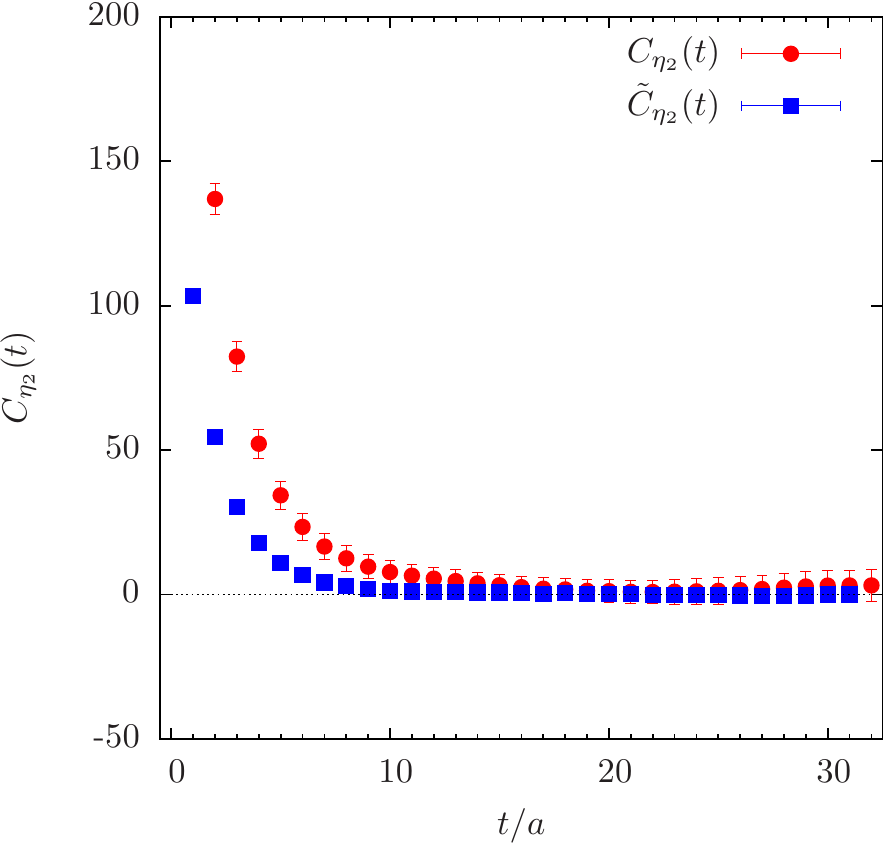}}
  \caption{$\eta_2$ correlation function $C_{\eta_2}$ and its shifted version $\tilde{C}_{\eta_2}$ as a function of $t/a$. For better visibility of the tail some of the numerically very large data points at small values of $t/a$ are not included in the plot. Left panel: Ensemble cA2.09.48. Right panel: Ensemble cA2.60.32.}
  \label{fig:Ceta2comp}
\end{figure}

In \fig{fig:Ceta2comp} we show $C_{\eta_2}$ and its shifted version
$\tilde{C}_{\eta_2}$ as functions of Euclidean time $t/a$, in the left
panel for the physical point ensemble cA2.09.48 and in the right panel
for cA2.60.32. For the physical point (left panel) we observe a sign
change in $C_{\eta_2}$ around $t/a=8$. However, from even slightly
earlier values of $t/a$, the correlation function is compatible with
zero, at least within two sigma. The point errors are large compared
to the observed fluctuations between different $t/a$ values
indicating large correlations. The shifting has two effects. First,
the error bars are dramatically decreased in $\tilde{C}_{\eta_2}$
compared to $C_{\eta_2}$ with at the same time strongly reduced
correlations. Second, $\tilde{C}_{\eta_2}$ turns negative only at
$t/a=18$ and stays compatible with zero within two sigma from then on.

In the right panel of \fig{fig:Ceta2comp} for $M_\pi\approx340\
\mathrm{MeV}$ the unshifted correlation function does not show a sign
change. Still, the shifted correlation function exhibits significantly
smaller error bars due to largely reduced correlations.

\begin{figure}[t]
  \centering
  \subfigure{\includegraphics[width=.49\linewidth]{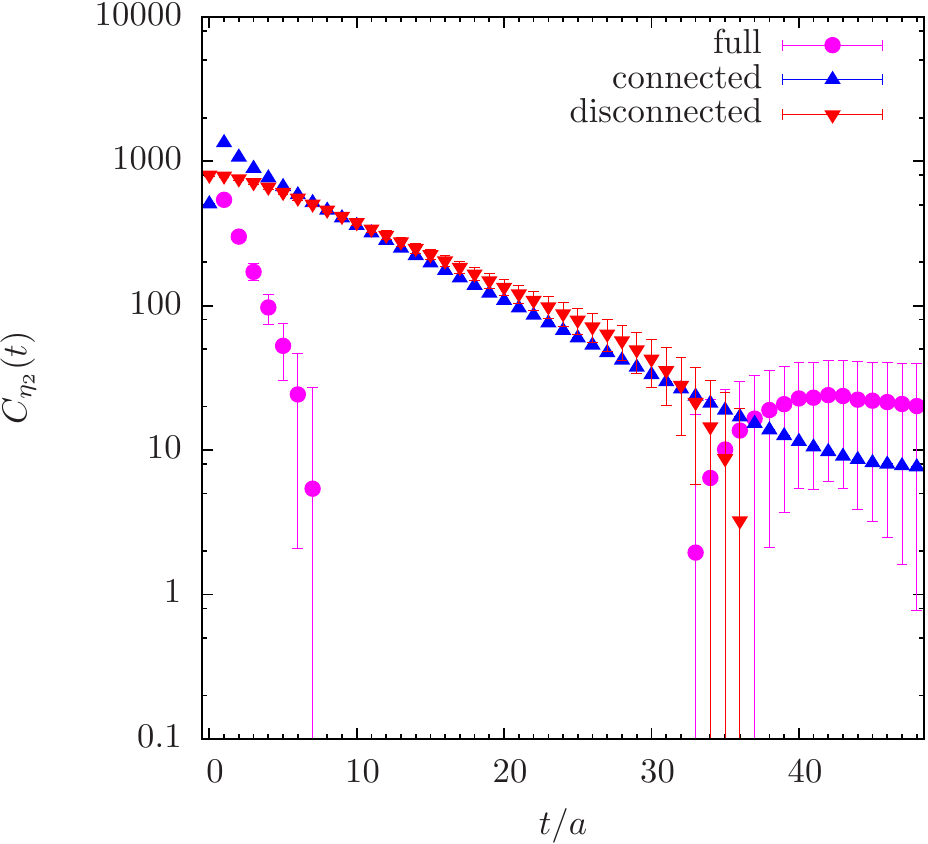}}
  \subfigure{\includegraphics[width=.49\linewidth]{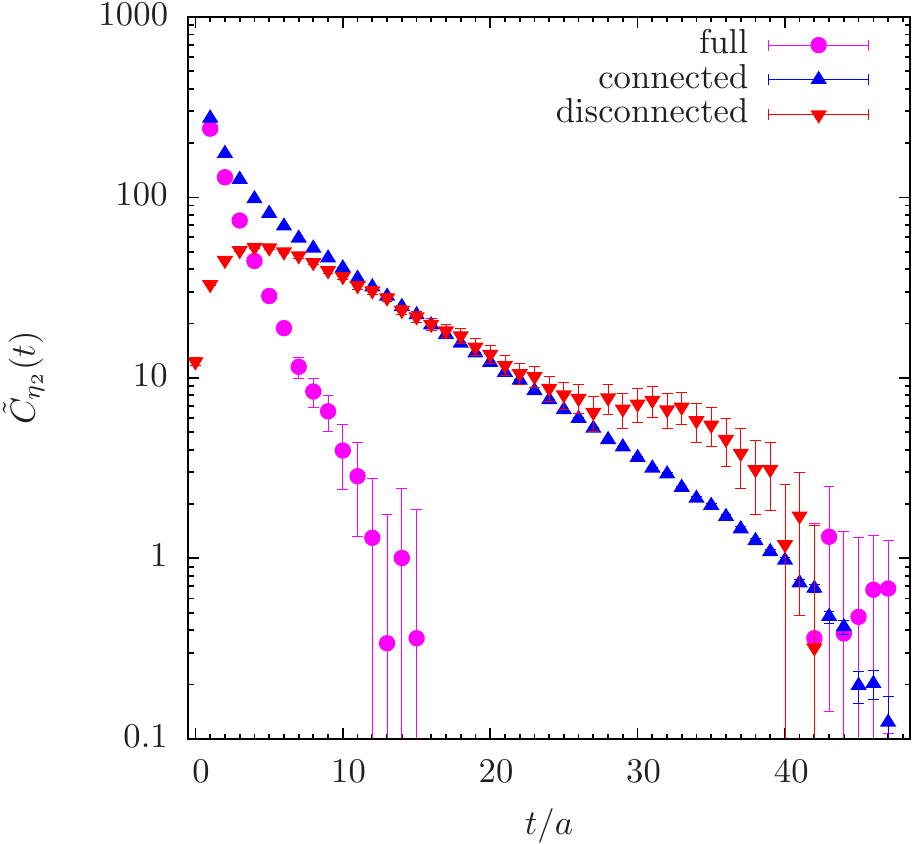}}
  \caption{Connected, disconnected and full $\eta_2$ correlation function versus $t/a$ for the physical point ensemble cA2.09.48. Left panel: Original correlation function $C_{\eta_2}$. Right panel: Shifted correlation function $\tilde{C}_{\eta_2}$.}
  \label{fig:cA2.09.48}
\end{figure}

In \fig{fig:cA2.09.48} we focus on the physical point ensemble
cA2.09.48. We show in a half logarithmic plot the connected,
disconnected and full $\eta_2$ correlation function versus $t/a$. We
recall that the full correlation function is obtained as the
difference between the connected and disconnected contribution,
cf.~\eq{eq:trcorr}.  In the left panel we show the unshifted
correlators and in the right panel the corresponding shifted
correlators. While the observations are the same as obtained from
\fig{fig:Ceta2comp}, the effect of the shift is better visible due to
the logarithmic scale on the $y$ axis.

Moreover, one sees from \fig{fig:cA2.09.48} that the signal-to-noise
ratio of the connected only contribution stays approximately constant
until close to $t=T/2$. Therefore, the connected correlation function
can be fitted at large Euclidean times using the ansatz
\begin{equation}
  \label{eq:fitansatz}
  f^\pm(t; A, M)\ =\ A\,\left(e^{-Mt} \pm e^{-M(T-t)}\right)\,,
\end{equation}
where the $\pm$ depends on whether the shifted or unshifted
correlation function is analyzed. Additionally, one learns from
\fig{fig:cA2.09.48} that the error on the full correlation functions
mainly stems from the disconnected contribution.

\begin{figure}[t]
  \centering
  \subfigure{\includegraphics[width=.49\linewidth]{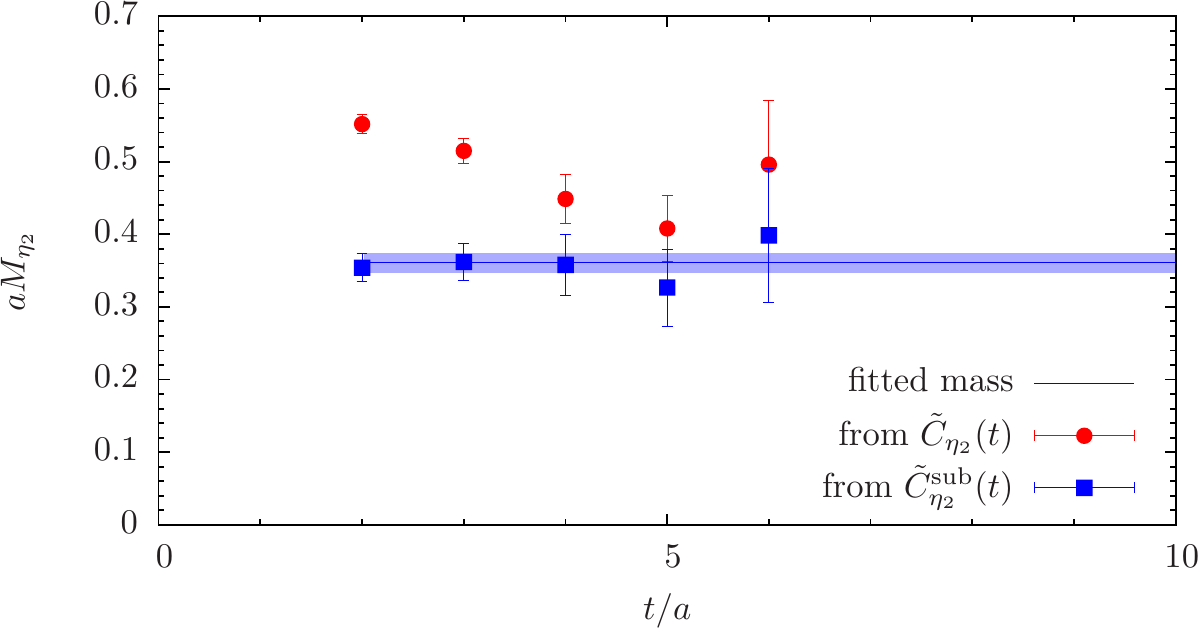}}
  \subfigure{\includegraphics[width=.49\linewidth]{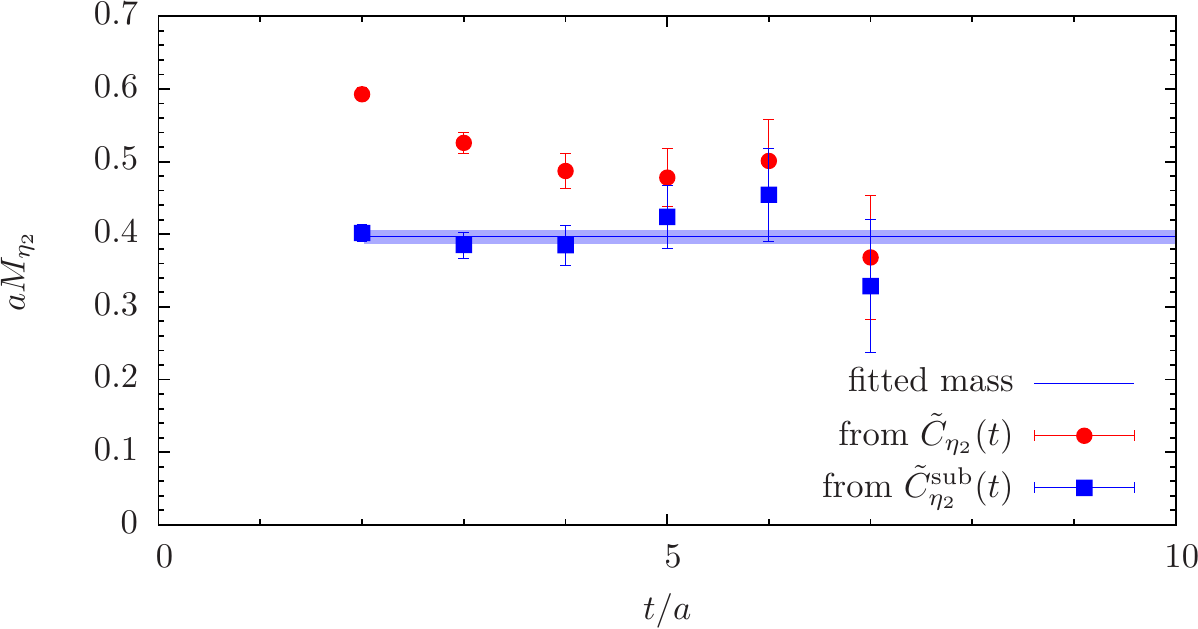}}
  \caption{Lattice data for effective masses computed from
    $\tilde{C}_{\eta_2}$ without and from
    $\tilde{C}_{\eta_2}^\mathrm{sub}$ with excited states
    subtracted. The result for the mass and its error from a
    correlated fit to the correlator data of
    $\tilde{C}_{\eta_2}^\mathrm{sub}$ has been included. Note that the
    end points of the fit ranges lie outside of the plots, because we
    fit to the correlation function and not to the effective
    masses. Left panel: Ensemble cA2.09.48. Right panel: Ensemble
    cA2.60.32.}
  \label{fig:Meffeta2comp}
\end{figure}

Once the connected-only part is fitted with the ansatz above, we can
apply the excited state subtraction as explained earlier. We denote
the corresponding subtracted and shifted $\eta_2$ correlation function
as $\tilde{C}^\mathrm{sub}_{\eta_2}(t)$. In \fig{fig:Meffeta2comp} we
show the effective masses computed from
$\tilde{C}^\mathrm{sub}_{\eta_2}(t)$ as a function of $t/a$. In the
left panel we show the data for the physical point ensemble cA2.09.48,
in the right panel for cA2.60.32. In both cases we observe a plateau
in the effective masses from $t/a=2$ or even $t/a=1$ on. The result of
a fit to the correlation function is indicated by the horizontal
lines, indicating also the fit range. The end points of the fit ranges
lie outside the plotted region, because we obtain a signal in the
correlation function further out in $t/a$.  For comparison, we also
show the effective masses computed from $\tilde{C}_{\eta_2}$ without
excited state subtraction, for which a plateau can clearly not be
identified with confidence.

The final values for $\meta$ are determined from a fit of ansatz
\eq{eq:fitansatz} to $\tilde{C}^\mathrm{sub}_{\eta_2}(t)$. The
corresponding results are compiled in \tab{tab:results}.

\subsection{\texorpdfstring{$\eta_2$}{eta2} Meson mass from topological charge density
  correlators}
\begin{figure}[thb] 
\subfigure{\includegraphics[height=.465\linewidth]{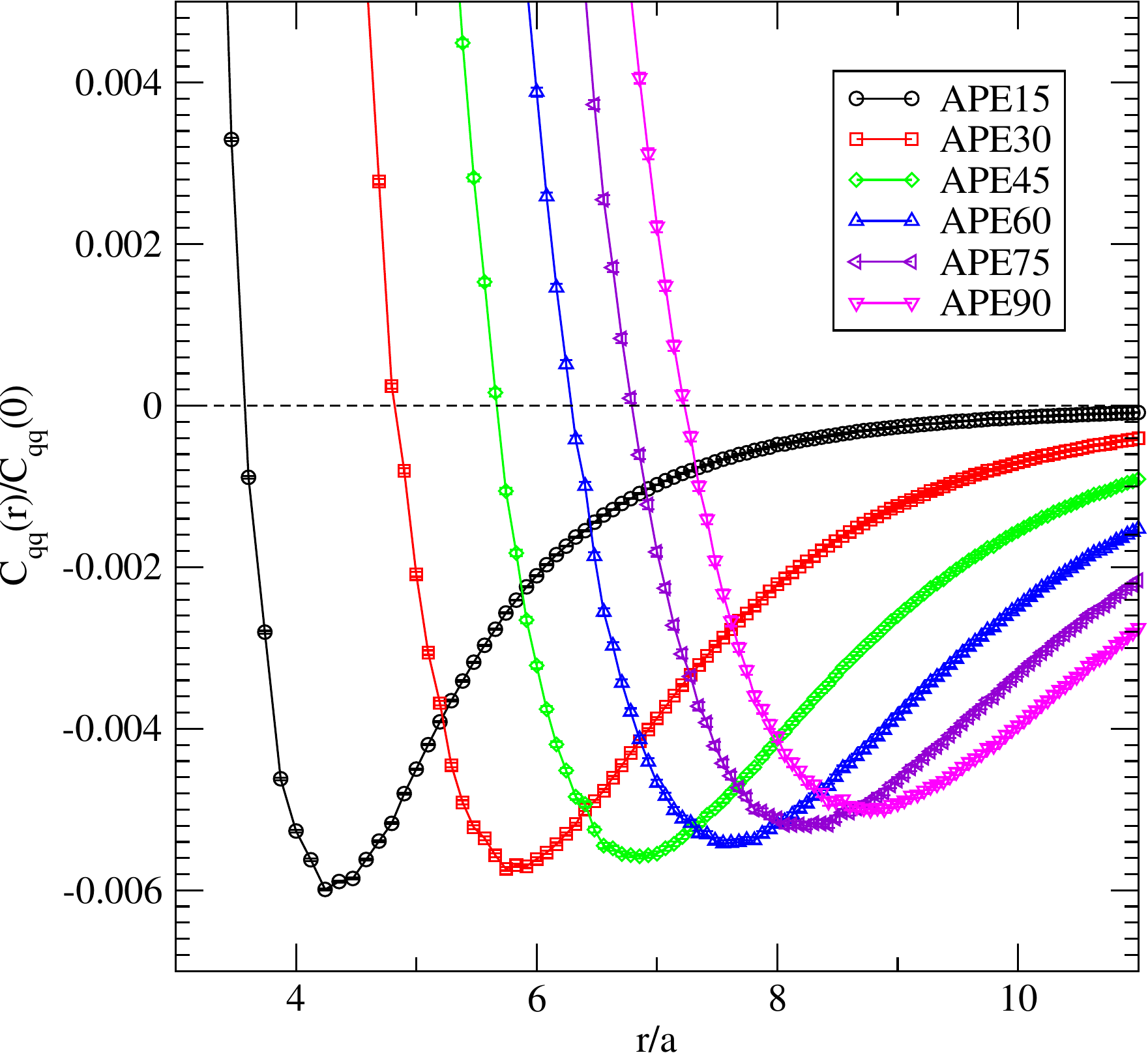}}
\subfigure{\includegraphics[height=.47\linewidth]{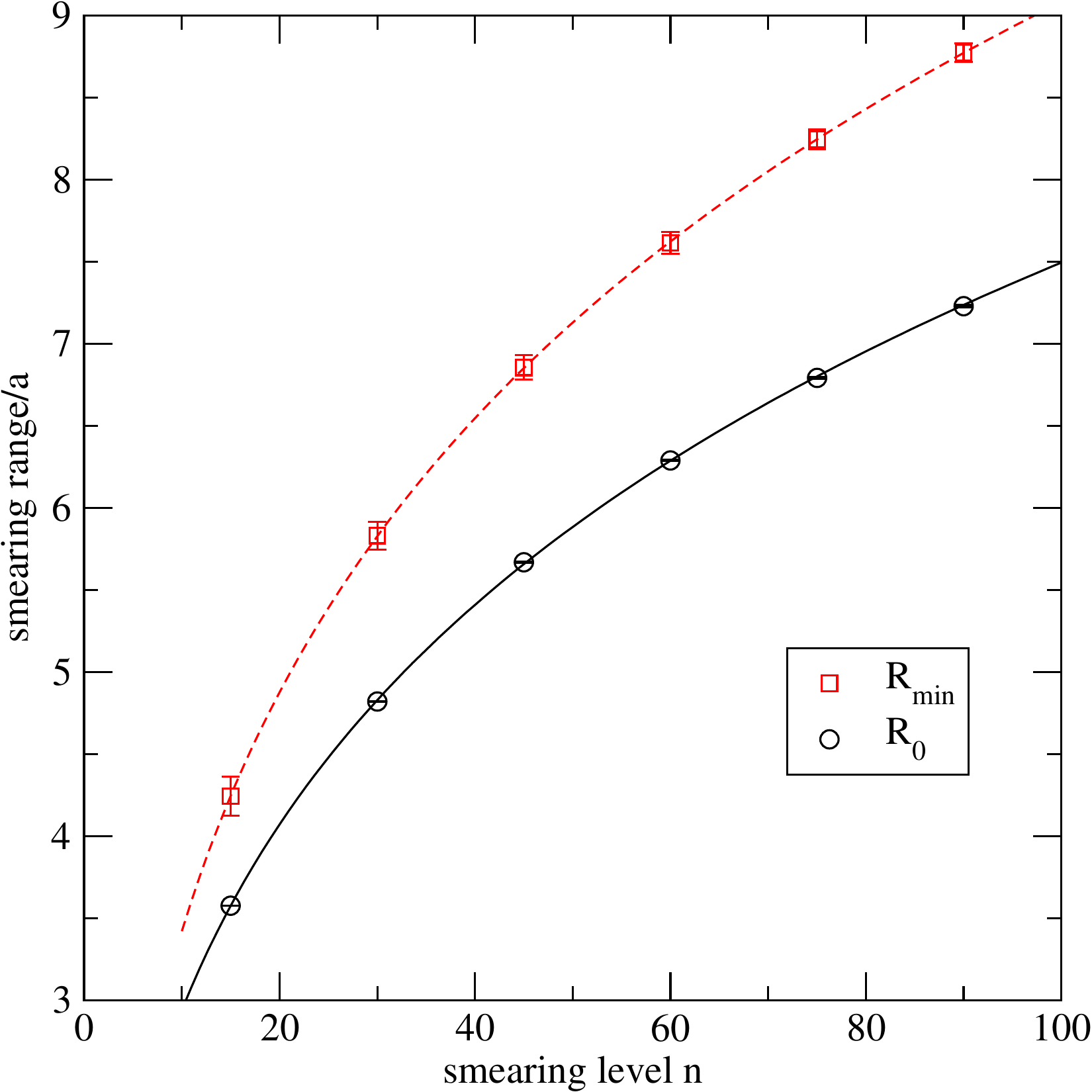}}
 \caption{Topological charge density correlator for various
    APE$n$ smearing with levels $n=15, 30, \ldots, 90$ for ensemble
    cA2.09.48. Left panel: Zoom of the normalized correlator 
    $C_{qq}(r)/C_{qq}(0)$ as a function of the separation $r$. Right panel: Scales characterizing the smearing range
    of the contact term as a function of the smearing levels.}
  \label{fig:ffd_correlator_APEXX_normalized_zoom}
\end{figure}

When determining the fit range in fitting the form in
\eq{eq:asymptotic_scalar_propagator} to the topological charge density
correlators $C_{qq}$, one needs to take into account the range over
which the contact term is smeared, as discussed above. For this
reason, we show in the left panel of
\fig{fig:ffd_correlator_APEXX_normalized_zoom} the correlators on
ensemble cA2.09.48 for different smearing levels $n=15,
30,45,60,75,90$. Since the maximum of the correlator at distance $r=0$
is suppressed with an increased smearing level and varies by an order of
magnitude between smearing levels $n=15$ and $n=90$, we normalize the
correlators by $C_{qq}(r=0)$.
The smearing range can be described by the two characteristic scales
$R_0$ and $R_\text{min}$, defined by the conditions $C_{qq}(r=R_0)=0$
and $C_{qq}(r=R_\text{min})$ where the correlator has its minimum
value. The dependence of these smearing ranges on the smearing levels
is displayed in the right panel of
\fig{fig:ffd_correlator_APEXX_normalized_zoom} for the correlators on
ensemble cA2.09.48 together with fits of the form $c_0+c_1 \log(n)+c_2
\log(n)^2$.

\begin{figure}[thb] 
\includegraphics[height=.465\linewidth]{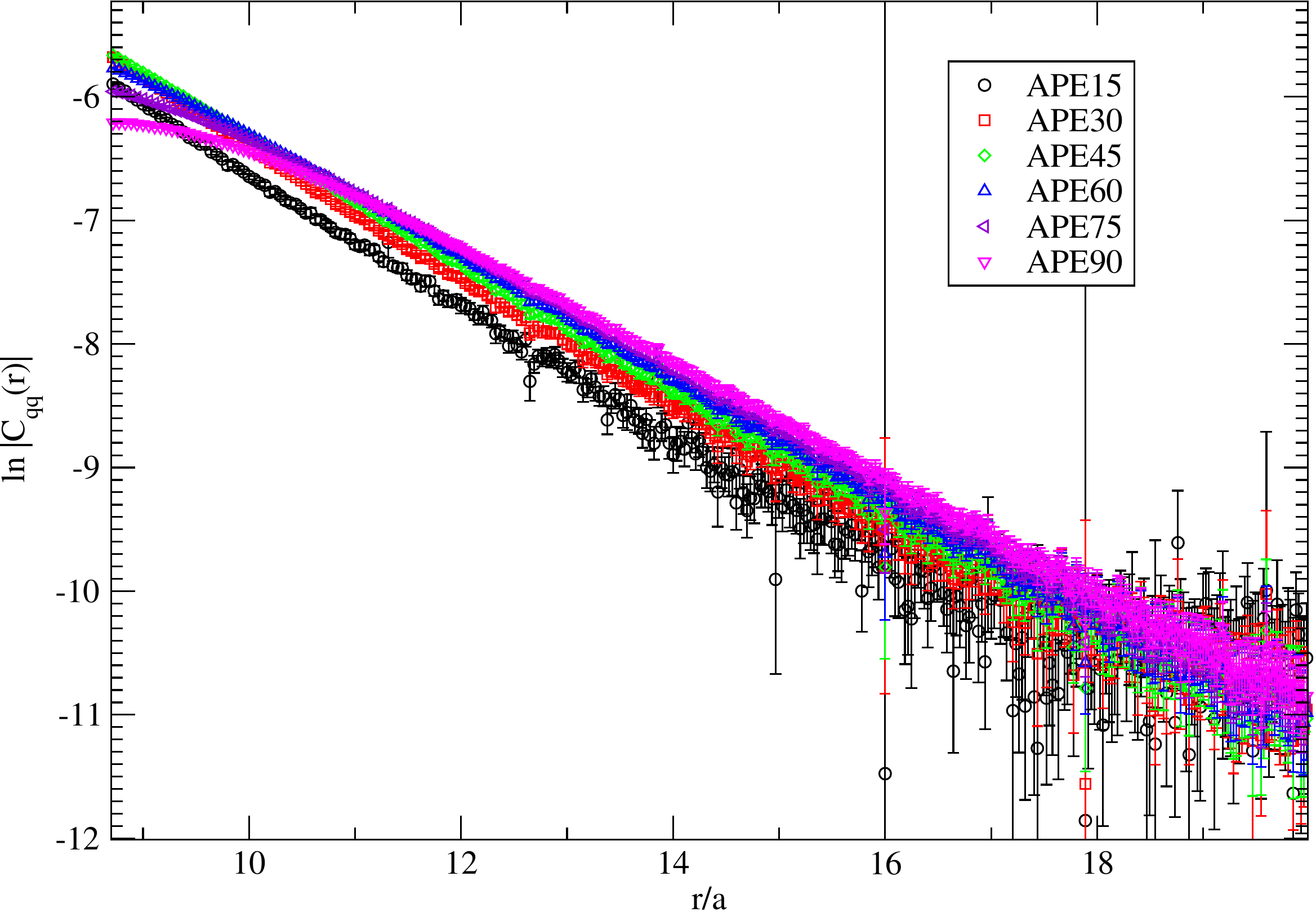}
 \caption{Long distance behavior of the topological charge density correlator for various
    APE$n$ smearing on ensemble
    cA2.09.48. }
  \label{fig:ffd_logcorrelator_APEXX}
\end{figure}

In \fig{fig:ffd_logcorrelator_APEXX} we show the long distance
behavior of the correlators on a logarithmic scale for the various
smearing levels. It is comforting to see that the correlators start to
asymptotically fall on top of each other for increasing smearing
level. Smearing levels $n=75$ and 90 for example are statistically
indistinguishable for $r/a \gtrsim 11$. Note that since the asymptotic
form of the correlator in Eq.~(\ref{eq:asymptotic_scalar_propagator}) is
\begin{equation}
C_{qq}(r) \sim \sqrt{\frac{M}{r}}\frac{1}{r} e^{-M r} \left(1 + {\cal O}\left(\frac{1}{Mr}\right)\right)
\quad \quad\text{for large $r$,}
\end{equation}
rather than purely exponential, the choice of the optimal fit range
cannot be guided by an effective mass plot. From 
\fig{fig:ffd_logcorrelator_APEXX} we infer that for the lowest
smearing level $n=15$ the signal is essentially lost after $r/a
\gtrsim 16$, while for the largest smearing level $n=90$ the fit range
is limited to $r\gtrsim 12$ due to the contamination by the
smeared-out contact term. Consequently, the intermediate smearing
levels seem to provide the longest fit ranges when both restrictions
are taken into account.

\begin{figure}[thb] 
\includegraphics[height=.465\linewidth]{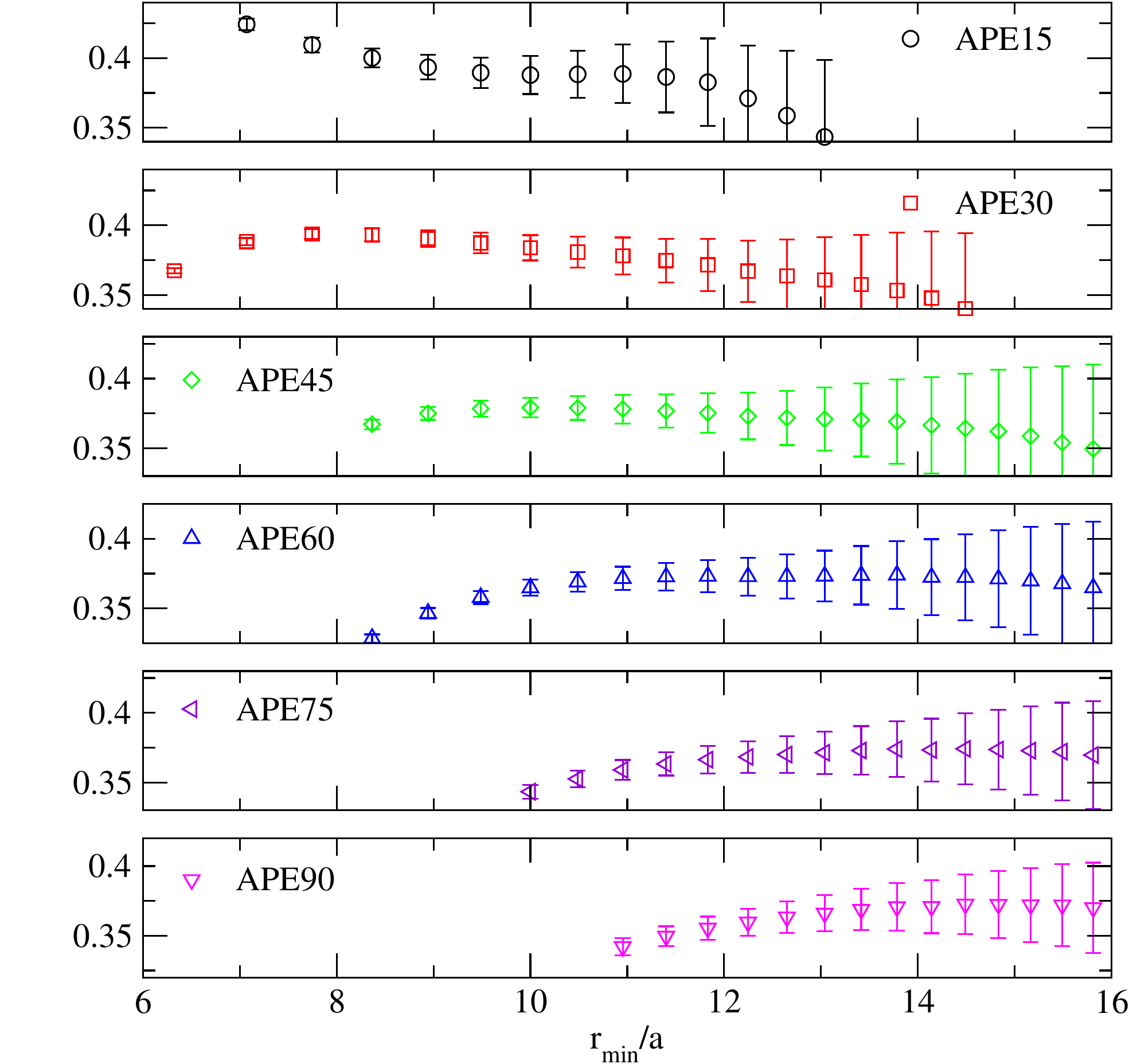}
\includegraphics[height=.465\linewidth]{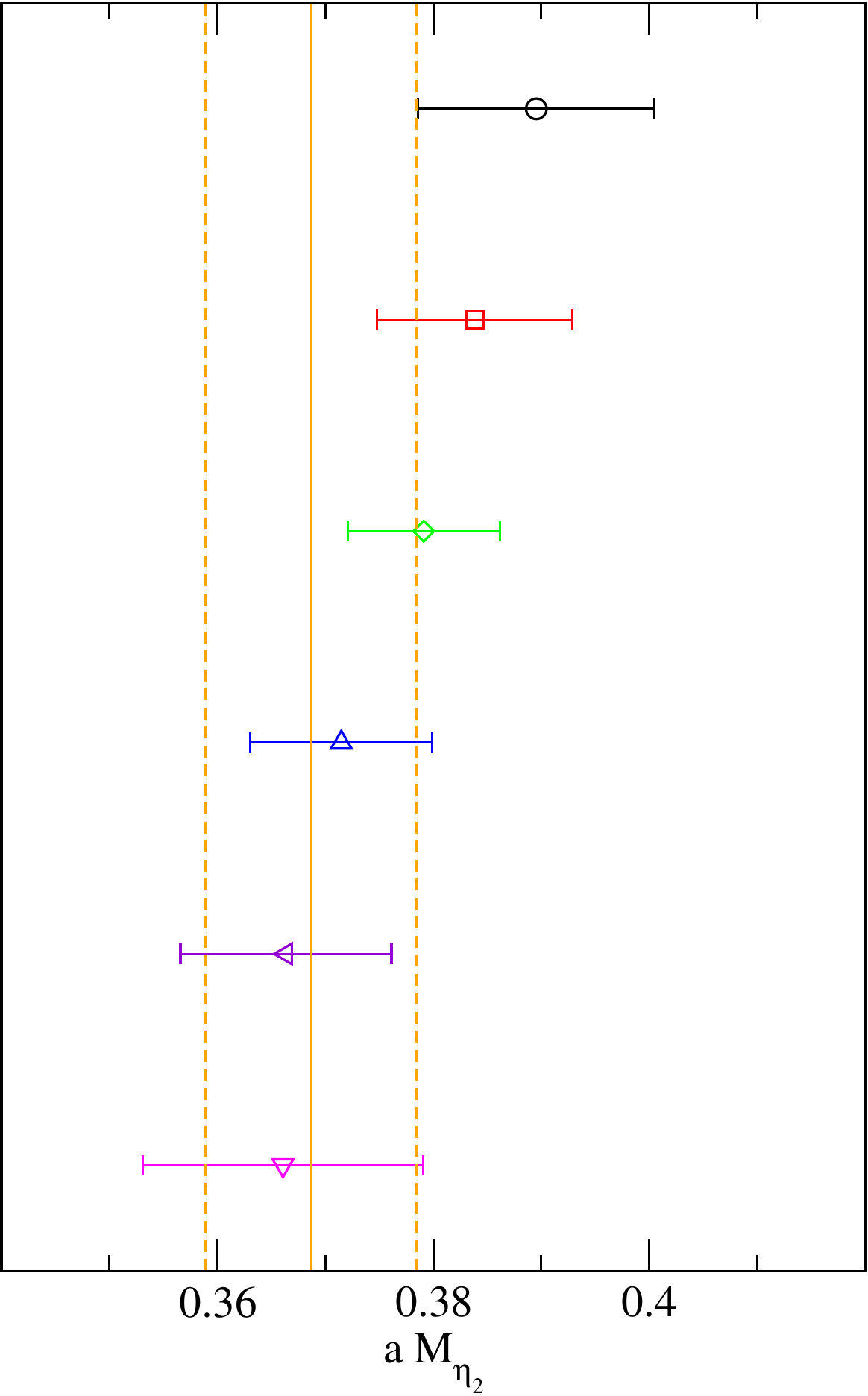}
 \caption{Fit results for different
    APE$n$ smearings on ensemble
    cA2.09.48. Left panel: As a function of $r_\text{min}/a$ for
    fixed $r_\text{max}/a=20$. Right panel: Variation of the fit result with the smearing level.}
  \label{fig:fitresults_vs_rmin_APEXX}
\end{figure}

\begin{figure}[thb] 
\includegraphics[height=.4\linewidth]{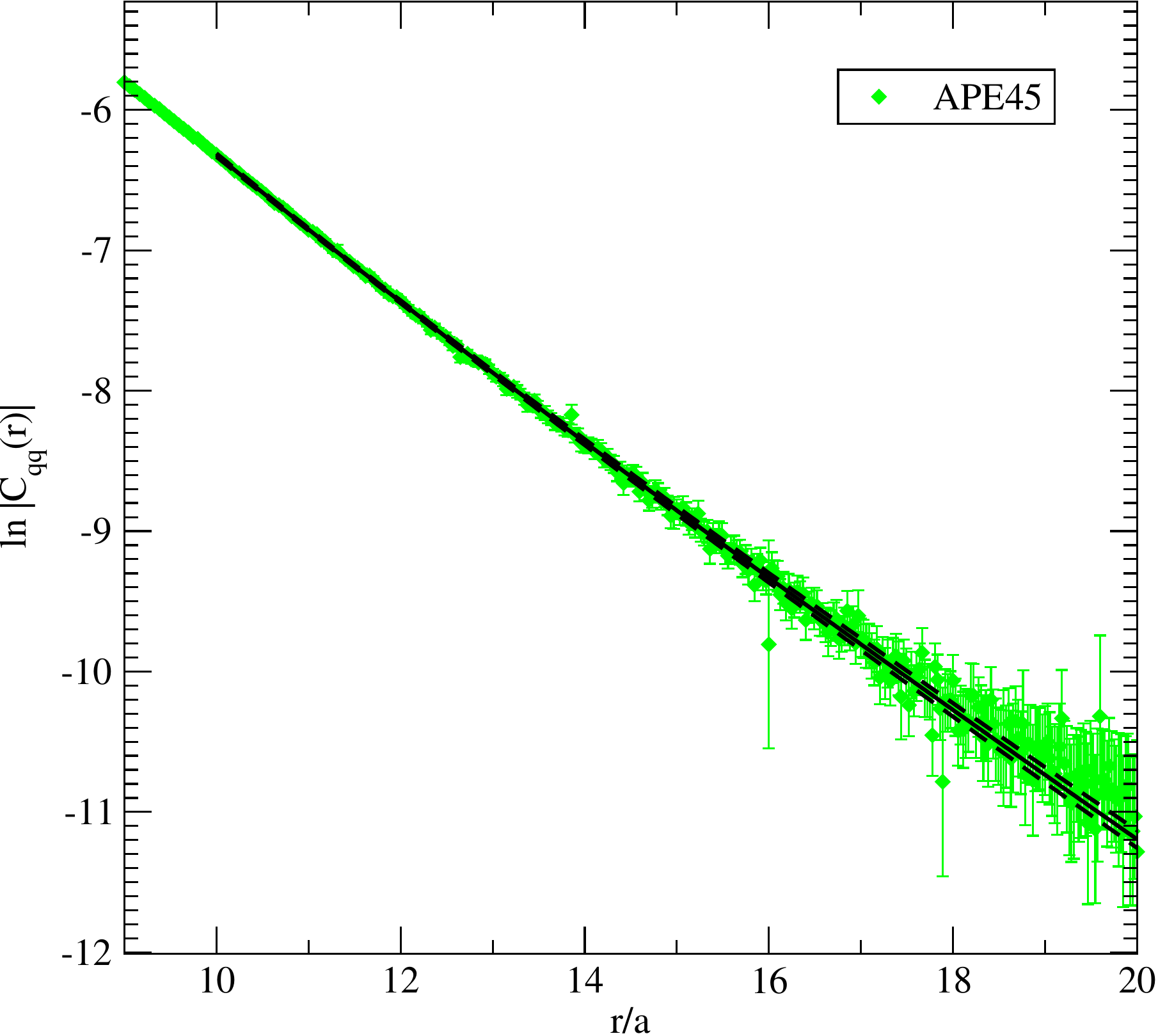}
\includegraphics[height=.4\linewidth]{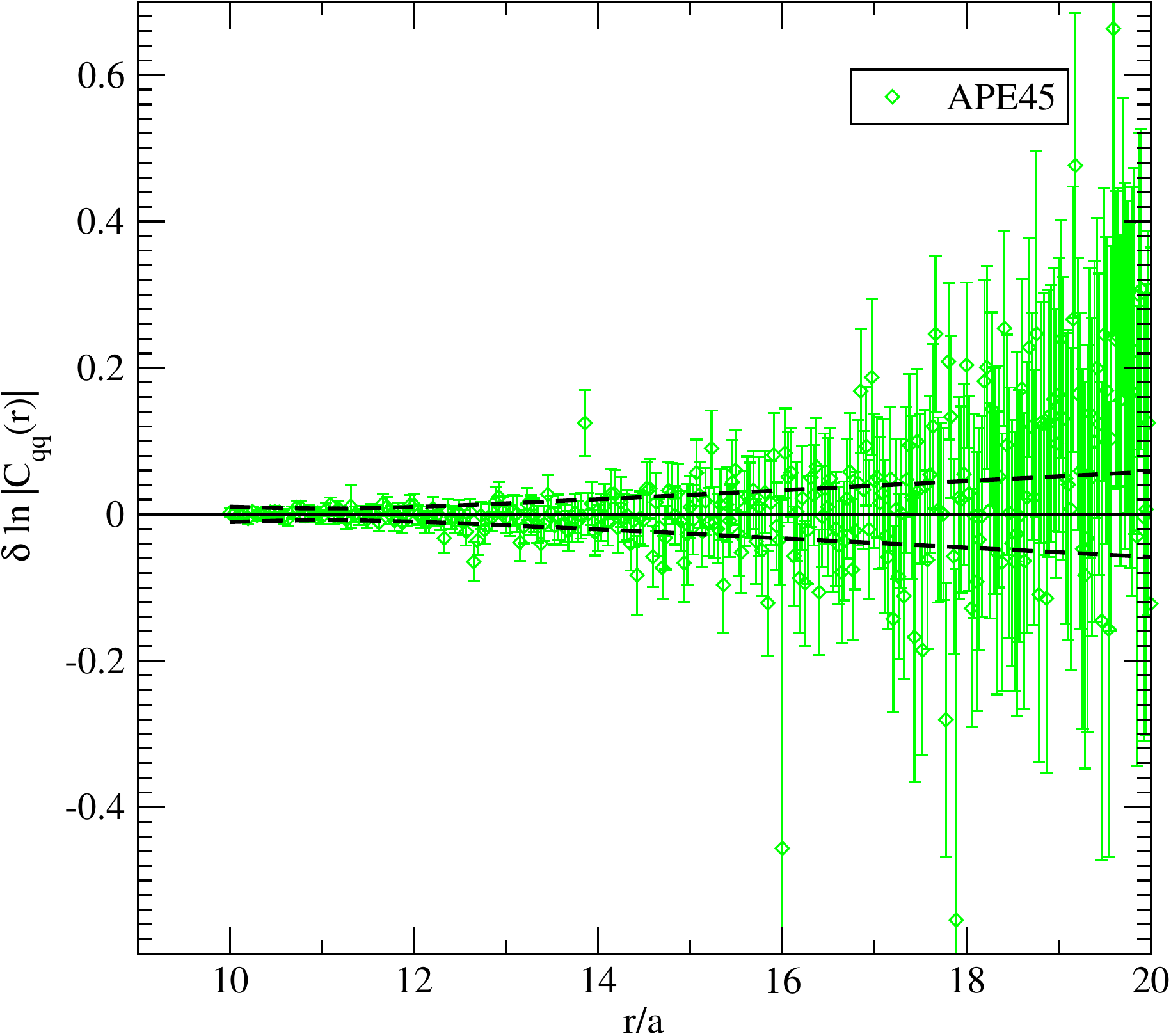}
 \caption{Example for a fit of the topological charge density
   correlation function with APE$45$ smearing on ensemble
    cA2.09.48. Left panel: Fit function and data. Right panel:
    Differences between fit function and data.}
  \label{fig:fit_ffd_logcorrelator_APE45}
\end{figure}

When trying to maximize the fit range $[r_\text{min},r_\text{max}]$
for the different smearing levels, we notice that the fit results
are not particularly sensitive to the choice of $r_\text{max}$ as long
as $r_\text{max} \gtrsim 16$. On the other hand, the error depends
strongly on the choice of $r_\text{min}$. This is illustrated in the left
panel of Fig.~\ref{fig:fitresults_vs_rmin_APEXX} where we show the
fit results for $aM$ as a function of $r_\text{min}/a$ while keeping
$r_\text{max}/a=20$ fixed. As we lower $r_\text{min}$ the error
becomes smaller, but at some point the fit result starts to change due to
the influence of the smeared contact term, and possibly also excited
state contributions. Consequently, for each smearing level we minimize
$r_\text{min}/a$ while making sure that the result is still stable
under a variation of $r_\text{min}/a$. In Fig.~\ref{fig:fit_ffd_logcorrelator_APE45} we give an
example for such a fit. The left panel shows the correlation function
on ensemble cA2.09.48 at smearing level $n=45$ together with the fit
function from a fit using $r_\text{min}/a=10$ and $r_\text{max}/a=20$,
while the right panel shows the differences between the fit function
and the data points. In this example we get $a M_{\eta_2}=0.3791(71)$
and $\chi^2/\text{dof}=0.61$ with 299 (correlated) degrees of
freedom. This result is very stable under a large variation of the fit
range.

Our choice for the $r_\text{min}/a$ values are $r_\text{min}/a \sim
9.5, 10.0, 10.0, 11.0, 12.0, 13.0$ for smearing level
$n=15,30,45,60,75,90$, respectively, and in the right panel of Fig.~\ref{fig:fitresults_vs_rmin_APEXX} 
we display the final fit result for each smearing level. 

Finally, we choose as our final value the weighted average between the
three smearing levels $n=60, 75$ and 90, at which the fit results seem
to stabilize, and we use the statistical error from the result at level
$n=75$ which also roughly covers the systematic error from varying
$n$. Our final result
\begin{equation}
a M_{\eta_2}^{\text{gl.}} =  0.3687(98) 
\end{equation}
is displayed in the right panel of Fig.~\ref{fig:fitresults_vs_rmin_APEXX} 
as the vertical orange band. We note that this is well compatible with the 
result from the fermionic correlators in Sec.~\ref{subsec:Results meson mass from fermionic correlators},
but it is here obtained from smeared topological charge density correlators 
which are significantly cheaper to calculate.

Repeating this procedure for the other ensembles yields the results
for $a M_{\eta_2}^{\text{gl.}}$ compiled in Table
\ref{tab:results}. We note that the values on the smaller lattice
volumes have a significantly larger error. This is mainly due to two
reasons. First and foremost, the calculations on the smaller lattices
cannot benefit from self-averaging as much as the ones on the larger
lattices. Second, due to the smaller lattice extent, the fitting
ranges, in particular $r_\text{max}$, are more restricted leading to a
larger variation of the fitted masses with the smearing levels and
hence to a larger systematic error.

\subsection{Topological susceptibility}

\begin{figure}[bth] 
  \centering
 \includegraphics[width=.6\linewidth,clip]{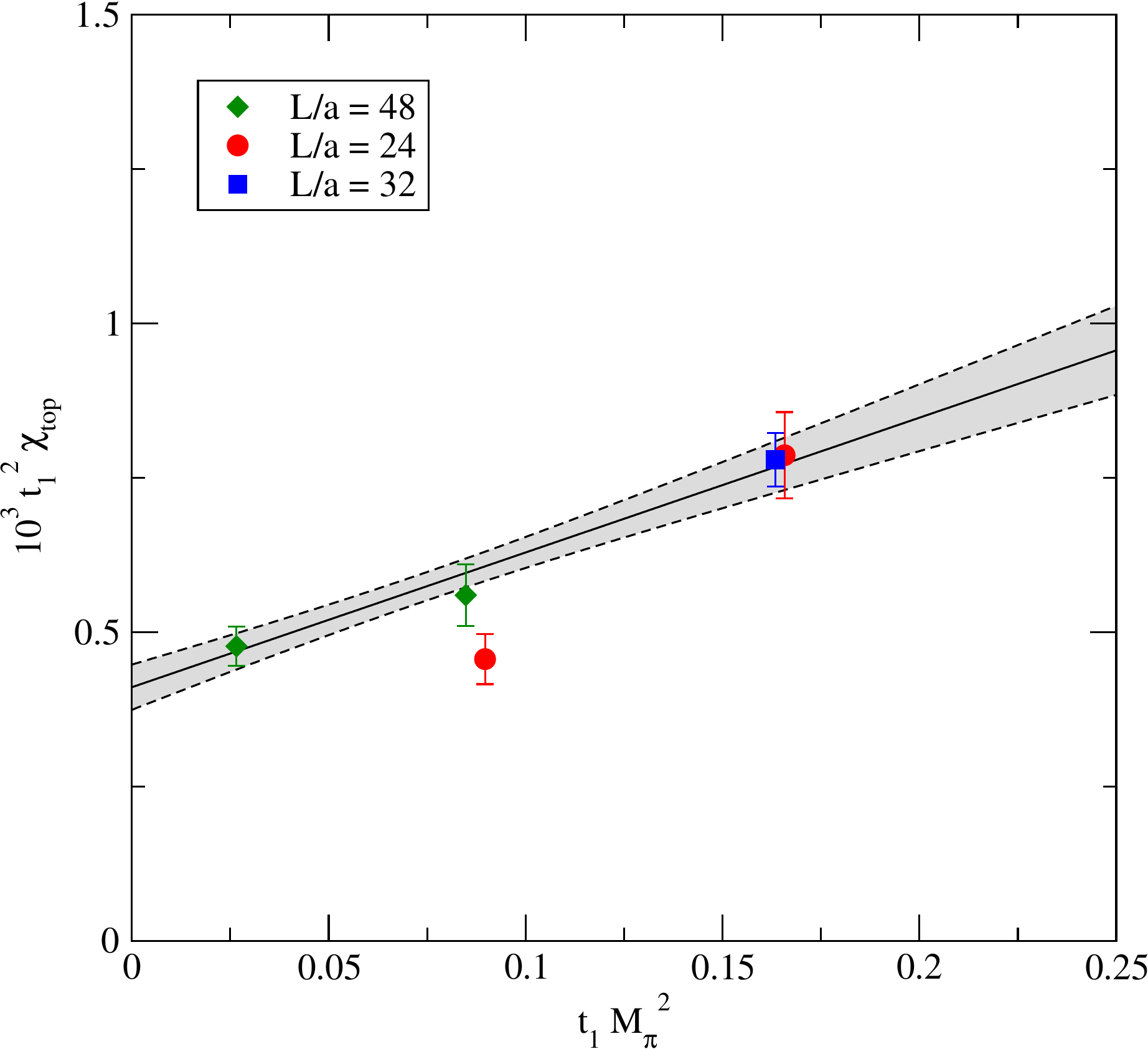}
  \caption{Topological susceptibility $\chi_\text{top}$ as a function of the
    squared pion mass, both in appropriate units of $t_1$. The solid
    line with shaded error band indicates a fit to the data
    according to \eq{eq:topsus}.}
  \label{fig:chit}
\end{figure}

In \tab{tab:results} we have also compiled our results for the
topological susceptibility evaluated at flow time $t=3 t_0$ as
discussed in Sec.~\ref{subsec:action_topcharge}, and the gradient flow
scale $t_1/a^2$. The values for $t_0/a^2$ can be found in Ref.~\cite{Abdel-Rehim:2015pwa}.  We express the susceptibility in units of $t_1$ in order to facilitate comparison with Ref.~\cite{Bruno:2014ova} and display the values in \fig{fig:chit} as a function of $t_1 \mpi^2$.
In leading order Wilson chiral perturbation theory one expects the following dependence of
$\chi_\text{top}$ on the lattice spacing and the pion mass~\cite{Bruno:2014ova}
written in units of the gradient flow scale $t_1$:
\begin{equation}
  \label{eq:topsus}
  t_1^2\, \chi_\text{top}\ =\ \frac{1}{8}\, t_1^2\, f_\pi^2\, \mpin^2\ +\ a^2\, \frac{c_2}{t_1}\,.
\end{equation}
Apart from the ensemble with too small volume cA2.30.24, our data are 
nicely compatible with this expectation: the solid line in
\fig{fig:chit} represents a fit of the function
\[
g(\mpi^2)\ =\ c_1 t_1 \mpi^2\ +\ a^2\, \frac{c_2}{t_1}
\]
to the data with fit parameters $c_1$ and $c_2$. Note that we use the
charged pion mass, because charged and neutral pion masses are
degenerate within errors. The best fit
parameter for $c_1$ is compatible with $t_1f_\pi^2/8$. Note that
ensemble cA2.30.24 has a very small volume explaining the outlier in
\fig{fig:chit}. 

The fitted value for $c_2$ can be compared to the results of
Ref.~\cite{Bruno:2014ova} using Wilson clover fermions. They obtain
$c_2 = 5.1(7)\times 10^{-3}$, while our value reads
$c_2=2.86(26)\times 10^{-3}$ indicating a sizable reduction of the
corresponding lattice artifact.

\section{Discussion}

\begin{figure}[t]
  \centering
  \includegraphics[width=1.\linewidth]{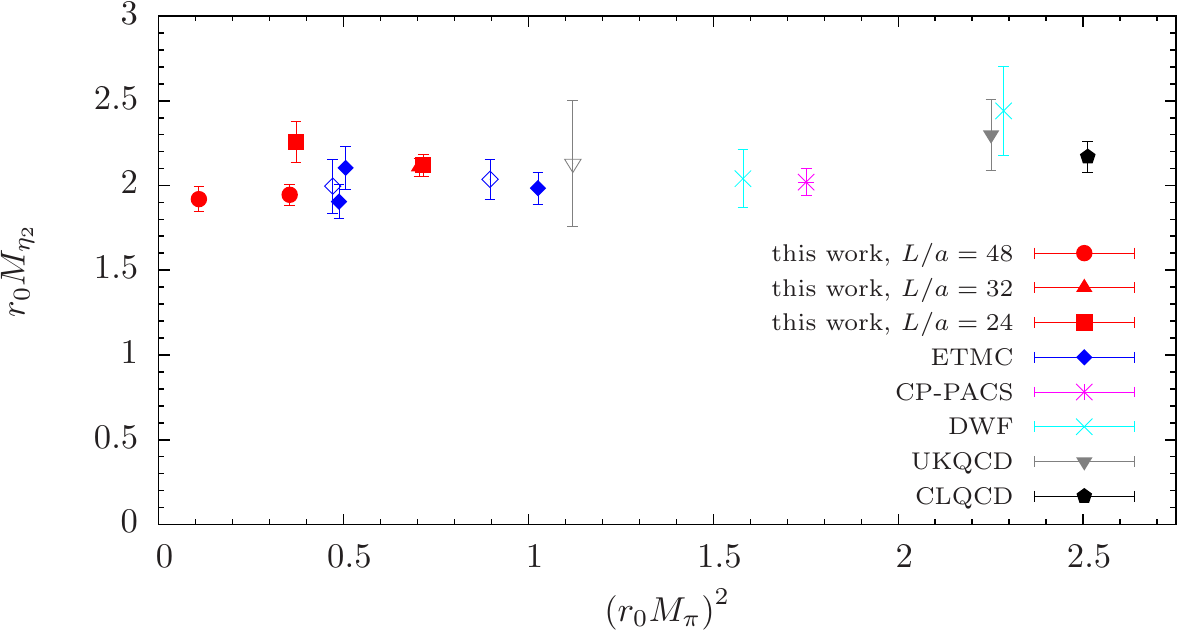}
  \caption{Compilation of literature values for the $N_f=2$
    $\eta^\prime$ meson: $r_0\meta$ as a function of
    $(r_0\mpi)^2$. The two UKQCD results stem from
    Ref.~\cite{Allton:2004qq} with the filled symbol for $r_0/a=5.04$
    and the open symbol for $r_0/a=5.32$, the PACS-CS result from
    Ref.~\cite{Lesk:2002gd} with $r_0/a=4.49$, the DWF result from
    Ref.~\cite{Hashimoto:2008xg} with $r_0/a=4.28$, the CLQCD result 
    from Ref.~\cite{Sun:2017ipk} with $r_0/a=4.22$ and the ETMC
    results from Ref.~\cite{Jansen:2008wv} with filled symbols for
    $r_0/a=5.22$ and open symbols for $r_0/a=6.61$.}
  \label{fig:r0eta2}
\end{figure}

In \fig{fig:r0eta2} we show $\meta^\mathrm{ferm.}$ in units of the
Sommer parameter $r_0$ as a function of $(r_0\mpi)^2$, with the value
of $r_0/a=5.317(48)$ from ensemble cA2.09.48 taken from
Ref.~\cite{Abdel-Rehim:2015pwa}. The outlier in our data points stems
again from the ensemble cA2.30.24, which has a very small value of
$M_\pi L$.  We compare the results presented in this paper determined
from the fermionic correlators to other lattice determinations
available in the literature: the two UKQCD results stem from
Refs.~\cite{Allton:2004qq}, the PACS-CS result from
Ref.~\cite{Lesk:2002gd} and the DWF result from
Ref.~\cite{Hashimoto:2008xg}. The twisted mass results without clover
are taken from Ref.~\cite{Jansen:2008wv}.

From this figure we conclude that there is overall very good agreement
between the different determinations. Even if the different
investigations do not cover a wide range in the lattice spacing, there
is no room for sizable lattice artifacts. The results presented in
this work complete the picture toward the physical point, with a
value
\[
r_0\meta\ =\ 1.92(8)
\]
at the physical point. Using $r_0=0.4907(86)$ from
Ref.~\cite{Abdel-Rehim:2015pwa} we arrive at  
\[
\meta\ =\ 772(18)\ \mathrm{MeV}
\]
where the scale setting error has been propagated into the final error
estimate. While the result is a bit lower than what is quoted in
Ref.~\cite{Jansen:2008wv}, the flat dependence of $\meta$ on the light
quark mass is confirmed.  Interestingly, this value agrees very well
with an estimate from Ref.~\cite{McNeile:2000hf}, where a
phenomenological analysis of the full $\eta, \eta^\prime$ mixing
matrix has been performed to arrive at $\meta\approx776\
\mathrm{MeV}$.

With this determination of $\meta$ at the physical pion mass value it
is almost certain that the $\eta_2$ meson will have a finite mass in
the chiral limit, agreeing with the picture that the $\eta_2$ is not a
Goldstone boson. It implies that the topological susceptibility must
decrease as $\mpi^2$ toward the chiral limit~\cite{Leutwyler:1992yt}.

\section{Summary}

In this paper we have presented results for the $\eta_2$ meson related
to the axial anomaly and the topological susceptibility in two-flavor
QCD. The results have been obtained using $N_f=2$ lattice QCD
ensembles generated by ETMC with the Wilson twisted clover
discretization~\cite{Abdel-Rehim:2015pwa}. Pion mass values reach from
the physical value up to $340\ \mathrm{MeV}$ at a single lattice
spacing value of $a=0.0931(2)\ \mathrm{fm}$. For the $\eta_2$ we could
confirm the almost constant extrapolation in $\mpi^2$ toward the
physical point. Errors are significantly reduced compared to previous
calculations. Lattice artifacts seem to be not larger than our
statistical uncertainty.

Regarding a future study of the $\eta$ and $\eta'$ at physical quark masses
in the $N_f=2+1+1$ theory we conclude that such a calculation should now be
feasible assuming a roughly similar signal-to-noise ratio as in the
two-flavor case. Since it is known from earlier $N_f=2+1+1$ simulations at
unphysical quark masses that the total error is dominated by the error on
the light quark disconnected loops, such an assumption seems reasonable.
While the nondegenerate heavy quark doublet will require additional
inversions, it should only lead to a moderate increase in the total
computational cost. An additional complication in the $N_f=2+1+1$ case
arises from the technically more involved analysis because --- unlike the
$\eta_2$ --- the $\eta'$ is not a ground state. However, all the relevant
analysis methods have been developed and successfully applied previously in
Refs.~\cite{Michael:2013gka,Ottnad:2017bjt} in a study of the $\eta$,
$\eta'$ at unphysical quark masses, and the analysis at physical quark masses
can be done in the same way.

We complement the determination of $\meta$ at the physical point from
fermionic correlation functions with one from the topological charge
density correlator. We find that with the number of APE smearing steps
larger than or equal to $60$ the estimated value of $\meta$ becomes
stable. The so determined value for $\meta$ is fully compatible with
the one from fermionic correlators and has an even smaller statistical
uncertainty. It is straightforward to apply this methodology in the
$N_f=2+1+1$ theory in order to determine the mass of the $\eta'$
meson: except for the mixing with the $\eta$, which can be taken into
account by appropriately modifying the fit function, we do not expect
any additional complications.

The topological susceptibility has been computed using the gradient
flow. As expected, $\chi_\text{top}$ is proportional to $\mpi^2$ (for small
$\mpi^2$) up to an additive lattice artifact independent of
$\mpi$. Even if we are not able to finally confirm this with only a
single lattice spacing at hand, this constant term should be of
$\mathcal{O}(a^2)$. The size of this artifact appears to be
significantly smaller than what is observed with Wilson clover
fermions in Ref.~\cite{Bruno:2014ova}.

\begin{acknowledgments}

We thank the members of ETMC for the most enjoyable
collaboration. We thank G.~Rossi for valuable comments on the
manuscript. 
The computer time for this project was made available
to us by the John von Neumann-Institute for Computing (NIC) on the
Jureca and Juqueen systems in J{\"u}lich and the HPC Cluster in
Bern.  This project was funded by the DFG as a project in the
Sino-German CRC110. U. W. acknowledges support from the Swiss National
Science Foundation.  The open source software packages
tmLQCD~\cite{Jansen:2009xp,Deuzeman:2013xaa,Abdel-Rehim:2013wba},
Lemon~\cite{Deuzeman:2011wz},
DD$\alpha$AMG~\cite{Alexandrou:2016izb}, and R~\cite{R:2005} have
been used.

\end{acknowledgments}

\bibliographystyle{h-physrev5}
\bibliography{bibliography}

\end{document}